\documentclass[twocolumn,superscriptaddress,amsmath,amssymb,pra]{revtex4-2}
\usepackage[utf8]{inputenc}
\usepackage[T1]{fontenc}
\usepackage{graphicx}
\usepackage{dcolumn}
\usepackage{appendix}
\usepackage{bm}
\usepackage{ulem}
\usepackage[usenames,dvipsnames]{xcolor}

\usepackage{subfigure}
\usepackage{bbm}
\usepackage{enumerate}

\usepackage[
  bookmarks=true,
  colorlinks,
  linkcolor=black,
  urlcolor=black,
  citecolor=black,
  plainpages=false,
  pdfpagelabels,
  final,
  breaklinks=true
]{hyperref}

\usepackage{titlesec}

\usepackage{array}

\usepackage{physics}

\newcommand{\opE}{\hat{\bf E}}
\newcommand{\opr}{\hat{\bf R}}
\newcommand{\meld}{{\bf d}}

\begin{document}
\title{Strong laser fields and their power to generate controllable high-photon-number coherent-state superpositions}

\author{J.~Rivera-Dean}
\affiliation{ICFO -- Institut de Ciencies Fotoniques, The Barcelona Institute of Science and Technology, 08860 Castelldefels (Barcelona)}

\author{Th. Lamprou}
 \affiliation{Foundation for Research and Technology-Hellas, Institute of Electronic Structure \& Laser, GR-7001 Heraklion (Crete), Greece}
\affiliation{Department of Physics, University of Crete, P.O. Box 2208, GR-71003 Heraklion (Crete), Greece}

\author{E. Pisanty}
\affiliation{ICFO -- Institut de Ciencies Fotoniques, The Barcelona Institute of Science and Technology, 08860 Castelldefels (Barcelona)}
\affiliation{Max Born Institute for Nonlinear Optics and Short Pulse Spectroscopy, Max Born Strasse 2a, D-12489 Berlin, Germany}
\affiliation{Department of Physics, King's College London, WC2R London, United Kingdom}

\author{P. Stammer}
\affiliation{ICFO -- Institut de Ciencies Fotoniques, The Barcelona Institute of Science and Technology, 08860 Castelldefels (Barcelona)}
\affiliation{Max Born Institute for Nonlinear Optics and Short Pulse Spectroscopy, Max Born Strasse 2a, D-12489 Berlin, Germany}

\author{A. F. Ordóñez}
\affiliation{ICFO -- Institut de Ciencies Fotoniques, The Barcelona Institute of Science and Technology, 08860 Castelldefels (Barcelona)}

\author{A. S. Maxwell}
\affiliation{ICFO -- Institut de Ciencies Fotoniques, The Barcelona Institute of Science and Technology, 08860 Castelldefels (Barcelona)}
\affiliation{Department of Physics and Astronomy, Aarhus University, DK-8000 Aarhus C, Denmark}

\author{M. F. Ciappina}
\affiliation{ICFO -- Institut de Ciencies Fotoniques, The Barcelona Institute of Science and Technology, 08860 Castelldefels (Barcelona)}
\affiliation{Physics Program, Guangdong Technion - Israel Institute of Technology, 241 Daxue Road, Shantou, Guangdong, China, 515063}
\affiliation{Technion -- Israel Institute of Technology, Haifa, 32000, Israel}

\author{M. Lewenstein}
\email{maciej.lewenstein@icfo.eu}
\affiliation{ICFO -- Institut de Ciencies Fotoniques, The Barcelona Institute of Science and Technology, 08860 Castelldefels (Barcelona)}
\affiliation{ICREA, Pg. Llu\'{\i}s Companys 23, 08010 Barcelona, Spain}

\author{P. Tzallas}
\email{ptzallas@iesl.forth.gr}
\affiliation{Foundation for Research and Technology-Hellas, Institute of Electronic Structure \& Laser, GR-7001 Heraklion (Crete), Greece}

\affiliation{ELI-ALPS, ELI-Hu Non-Profit Ltd., Dugonics tér 13, H-6720 Szeged, Hungary}

\date{\today}

\begin{abstract}
    Recently, intensely driven laser-matter interactions have been used to connect the fields of strong laser field physics with quantum optics by generating non-classical states of light. Here, we make a further key step and show the potential of strong laser fields for generating controllable high-photon-number coherent-state superpositions. This has been achieved by using two of the most prominent strong-laser induced processes: high-harmonic generation and above-threshold ionization. We show how the obtained coherent-state superpositions change from an optical Schrödinger ``cat'' state to a ``kitten'' state by changing the atomic density in the laser-atom interaction region, and we demonstrate the generation of a 9-photon shifted optical ``cat'' state which, to our knowledge, is the highest photon number optical ``cat'' state experimentally reported. Our findings anticipate the development of new methods that naturally lead to the creation of high-photon-number controllable coherent-state superpositions, advancing investigations in quantum technology.
\end{abstract}

\maketitle

\section{INTRODUCTION}
Strong laser field physics and quantum optics are two research directions founded on the classical and quantum description of the electromagnetic field, respectively. Quantum optics has proven to be a very important field towards the development of quantum technologies \cite{Acin,Walmsay,Deutsch}, advancing studies ranging from fundamental tests of quantum theory to quantum information processing and quantum communication protocols. Central to these applications lies the concept of non-classical light states, that is, states of light that can be described only in a quantum mechanical frame \cite{VogelWelsch,QV2,QV3}. Within the family of non-classical light states, the superposition of two distinct coherent-states, i.e. the so-called optical \emph{Schrödinger cat states}, have proven to be a potentially useful candidate for the aforementioned applications \cite{Lvovsky2020,Ralph2003,Sanders1992,Jeong2003,Stobinska2007,Munro2002}. However, despite the progress there has been so far towards their practical generation \cite{Zavatta2004, Dakna1997,Ourjoumtsev2006,Ourjoumtsev2007,Hacker2019, Sychev2017}, the applicability of the existing optical cat states is partially restricted by their low photon number. Furthermore, the development of new schemes for the generation of high-photon-number optical cat states with controllable quantum features is considered a challenging task.

On the other hand, strong laser field physics \cite{Mourou, Strickland,Corkum1993,Lewenstein1994,Symphony,Kulander1993,SalieresScience} is a widely active research direction which has opened the way for studies ranging from relativistic electron acceleration (see \cite{Mourou} and references therein) to ultrafast electronics (see \cite{Vampa2018,Ciappina,Kruchinin2018} and references therein). Central to these investigations is the interaction of atoms with intense laser fields, which leads to the generation of coherent radiation in the extreme ultraviolet (XUV) \cite{McPherson1987, Ferray1988, AnneML, Hergott, Constant, Heyl, Symphony} and the X-ray \cite{Jens,Murnane} regimes, and has been substantially applied in attosecond science \cite{Symphony, Krausz, Ciappina}, non-linear XUV optics \cite{Kobayashi, Midorikawa, Chatziathanasiou, Tsatrafyllis, Boris, Nayak, Vrakking, Orfanos}, high resolution spectroscopy \cite{Gohle, Cingo} and tomography \cite{Young, Paulus}. The majority of these studies are experimentally conducted using high power femtosecond laser sources, and its interaction with matter is theoretically described by approaches where the electromagnetic field is treated classically.

Despite the large progress achieved in quantum optics and strong laser field physics, the direction of both research domains has remained uncoupled over the years. This is primarily due to the highly successful treatment of a classical electromagnetic field in strong laser physics and the assumption that the quantum aspects of the field were superfluous. Thus, the advantages emerging from the connection between quantum optics and strong laser field physics remain largely unexploited. However, very recently a link between both disciplines has been achieved theoretically and experimentally by showing that intense laser-matter interactions can lead to the generation of optical Schrödinger cat states \cite{Lewenstein1}.

Here, we make a key step forwards and show the power of the strong laser fields for the generation of controllable high-photon-number coherent-state superpositions. This has been achieved using the processes of high-harmonic generation (HHG) and above-threshold ionization (ATI) induced in intense laser-atom interactions. Specifically, we study the back-action of these two processes on the initial coherent-state of the driving field, analyze its phase space dynamics within a cycle of the field and along the duration of the driving pulse envelope, and show how the key action of conditioning on HHG and ATI processes can naturally lead to the generation of coherent-state superpositions of arbitrary high-photon-number. We also discuss how the laser-atom interaction conditions, in experiment, can be used to control the quantum features of these states. The theoretical results have been confirmed experimentally by showing the dependence of the non-classical features of the generated light after HHG on the atomic gas pressure. Furthermore, to demonstrate the high-photon nature of the generated cat states, we have experimentally achieved a 9-photon number optical Schrödinger cat state.

The paper is organized as follows. In Sec.~\ref{Th:background} we present the Schrödinger equation governing the interaction between the atom and the quantized field and, in Secs.~\ref{Section:HHG} and \ref{Section:ATI}, we condition the obtained equation to the HHG and ATI processes and study the obtained quantum optical states. In Sec.~\ref{Sect:ExperimentalSetup} we describe the experimental setup that allows for the conditioning onto HHG, which is later used in Sec.~\ref{Section:density} to show the transition from a ``cat'' state to a ``kitten'' by changing the atomic density in the interaction region, and in Sec.~\ref{Section:HighCats} to show the high-photon-number nature of the obtained quantum optical cat states. In the present work, we consider as low-- and high--photon number states, the states having mean photon number $\langle n \rangle$ in the range $\langle n \rangle \leq 2$ and $\langle n \rangle > 5$, respectively. Finally, in Sec.~\ref{Discussion} we provide a discussion on the perspective towards future implications of this work.

\section{THEORY}
\begin{figure*}[ht!]
    \centering
    \includegraphics[width=1\textwidth]{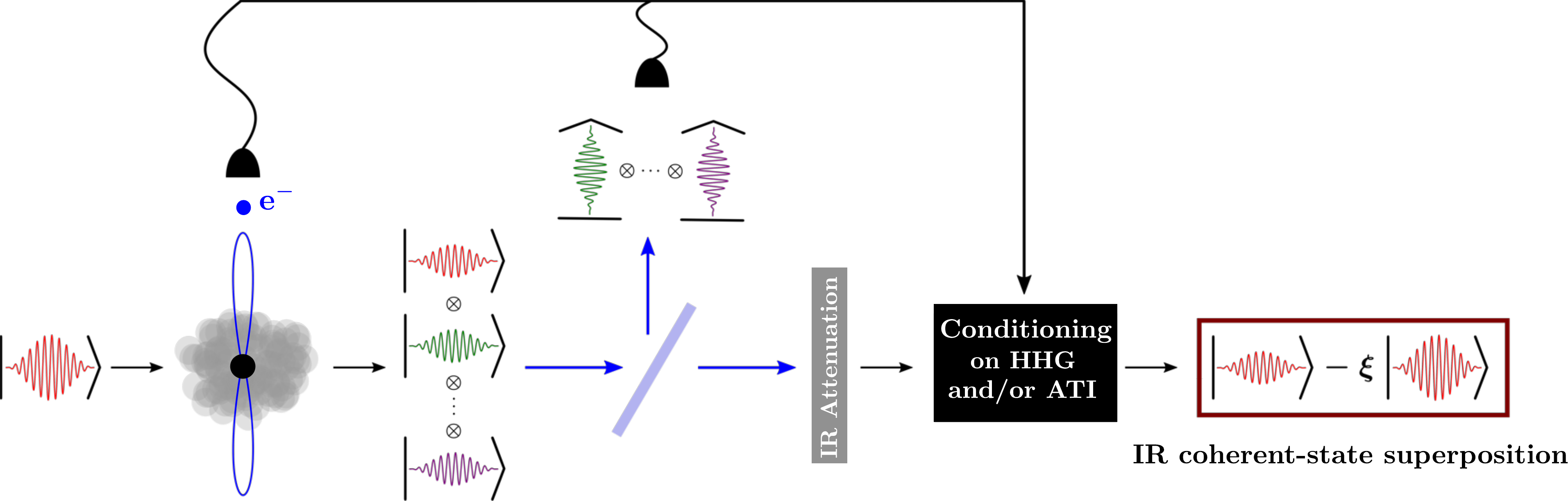}
    \caption{Scheme of the different \emph{conditionings}. A high-photon-number coherent-state coming from a laser source interacts with an atomic gas jet. As a consequence of the strong-field interaction that takes place, some electrons will ionize and, subsequently, may either recombine with the parent ion generating high-order harmonics or stay in the continuum. Thus, depending on the particular process we want to consider, we can look at the generated harmonics in case we want to study the quantum optical state of light obtained after HHG, and/or we can measure the generated photoelectrons in order to include ATI processes. We can further constraint this last measurement to photoelectrons that have a specific kinetic momentum, or consider all possible momenta. As a consequence of these conditioning measurements, the final quantum optical state of the IR mode can be written as superposition two or more coherent-states.}
    \label{Fig1}
\end{figure*}

\subsection{Theoretical background}\label{Th:background}

The qualitative understanding of the interaction is traditionally provided by the well-known three-step model \cite{Corkum1993, Kulander1993, Lewenstein1994}. According to this model, when a low frequency (usually in the infrared (IR) spectral range) intense linearly polarized laser field interacts with an atom or molecule, an electron tunnels out from the considered system, then it accelerates in the continuum gaining energy from the laser field and, within the same cycle of the field, it may re-collide elastically or inelastically with the parent ion. This process is repeated every half cycle of the laser field leading to the generation of ions, photoelectrons or photons with frequencies higher than the driving laser field (high harmonics (HH)). The non-recolliding electrons and the electrons that re-collide elastically with the parent ion contribute to the generation of above-threshold ionization photoelectrons \cite{PaulusATI, AnneML}, while the inelastic recollision leads to the generation of HH (electron recombines with the ion emitting a photon) or multiple charged ions (for example, via non-sequential double ionization) \cite{AnneML}.

Our fully quantized theoretical approach relies on the study of the reduction of the amplitude ($\delta \alpha_L$) in the initial coherent-state of the fundamental mode ($\ket{\alpha_L}$) as a consequence of its interaction with the atomic ensemble. The performance of further quantum operations, which we shall refer hereupon to as \emph{conditionings}, allow us to constrain our equations to specific strong-field physics processes, in particular to HHG and ATI. As a consequence of these operations, schematically illustrated in Fig.~\ref{Fig1}, the outgoing final state of the fundamental mode is given as the superposition of amplitude shifted coherent-states, as it was shown in  \cite{Lewenstein1} for the HHG scenario.

Briefly (further details about the calculations can be found in \ref{SM1}), we start from the time-dependent Schrödinger equation (TDSE) describing the interaction of the quantized field with the considered atom within a single active electron picture and in the dipole approximation. After performing a set of unitary transformations, it can be shown that the final TDSE characterizing the joint state between the electron and the field modes is given by
\begin{equation}\label{eq:total:TDSE}
    i \hbar \pdv{}{t}\ket{\psi(t)} 
        = - \text{e}\opE_Q(t) \cdot \opr_H(t) \ket{\psi(t)},
\end{equation}
where $\text{e}\opr_H(t)$ is the time-dependent dipole operator in the so-called semiclassical interaction picture (see \ref{SM1}) acting exclusively on the electronic degrees of freedom, and $\opE_Q(t)$ is a discrete version the electric field operator acting on the fundamental modes and its harmonics up to the cut-off region of the spectrum, that is,
\begin{equation}\label{eq:Efield}
    \opE_Q(t) = -i\hbar \vb{g}(\omega_L)f(t)
                \Big[
                    \big(
                        \hat{a}^\dagger - \hat{a}
                    \big)
                    + \sum_{q=2}^\text{cutoff} 
                        \sqrt{q}\big(
                                    \hat{b}_q^\dagger - \hat{b}_q
                                \big)
                \Big],
\end{equation}
where $\hat{a}$ ($\hat{a}^\dagger$) and $\hat{b}_q$ ($\hat{b}_q^\dagger$) are the annihilation (creation) operators acting on the fundamental and $q$-th harmonic respectively, $\vb{g}(\omega_L)\propto \sqrt{\omega_L/V_\text{eff}}$ is the coefficient that enters into the expansion of the laser electric modes and that depends on $V_\text{eff}$ which is the effective quantization volume \cite{QV1, QV2}, and $0 \leq f(t) \leq 1$ is a dimensionless function describing the pulse envelope.

Note that, as a consequence of the intense laser-atom interaction, we can either find the electron remaining in the ground state or in a continuum state. In the following, we show how Eq.~\eqref{eq:total:TDSE} can be used for the quantum optical description of two of the most central processes in strong-field physics: HHG and ATI.

\subsection{Quantum optical dynamics of HHG}\label{Section:HHG}
In the HHG process, the electron gets first transferred to the continuum via tunneling ionization due to the strong laser field we are applying and, later on, it recombines with the parent ion that was left behind, ending up again in the ground state of the system. Therefore, in order to get information about the HHG photonic quantum state, we condition Eq.~\eqref{eq:total:TDSE} onto the atomic ground state $\ket{\text{g}}$, i.e.,
\begin{equation}\label{eq:HHG:TDSE}
	i \hbar \pdv{}{t} \braket{\text{g}}{\psi(t)}
		= -\opE_Q(t) \cdot \mel{\text{g}}{\text{e}\opr_H(t)}{\psi(t)}.
\end{equation}

After strong-field physics approximations, the above equation can be expressed as (see \ref{SM2})
\begin{equation}
    i\hbar \pdv{}{t}\ket{\Phi(t)}
        = - \opE_Q(t) \cdot \meld_H(t) \ket{\Phi(t)},
\end{equation}
where $\ket{\Phi(t)} = \braket{\text{g}}{\psi(t)}$ and $\meld_H(t) = \mel{\text{g}}{\text{e}\opr_H(t)}{\text{g}}$ is the averaged time-dependent dipole operator. Here, $\meld_H(t)$ can be easily computed by numerically solving the TDSE, or by means of the strong-field approximation (SFA) theory \cite{Symphony, Lewenstein1994, ScrinziBook}. Whatever the method used, this equation can be easily solved as it is written as a linear combination of photon creation and annihilation operators for the different modes considered in the problem. This has a natural implication, that the final solution is given by a product state of all the modes participating in the process,
\begin{equation}\label{eq:HHG:qstate}
	\begin{aligned}
	\ket{\Phi(t)}
		= \ &e^{i\varphi_L(t)}\ket{(\alpha_L + \delta\alpha_L)e^{-i\omega_L t}}
			\otimes e^{i\varphi_2(t)}\ket{\beta_2 e^{-i2\omega_L t}}\\
			&\otimes \dots
			\otimes e^{i\varphi_q(t)}\ket{\beta_q e^{-iq\omega_L t}}
			\otimes \dots,
	\end{aligned}
\end{equation}
where $\delta \alpha_L(t)$ and $\beta_q(t)$ are defined as
\begin{equation}\label{eq:def:deltaalpha}
	\delta \alpha_L(t)
		= N {\bf g}(\omega_L) \cdot 
          \int^t_{t_0}\dd \tau\ 
            f(\tau) {\bf d}_H(\tau) e^{i\omega_L \tau}
\end{equation} 
\begin{equation}\label{eq:def:beta}
	\beta_q(t)
		= N \sqrt{q}\ {\bf g}(\omega_L) \cdot 
          \int^t_{t_0}\dd \tau\ 
            f(\tau) {\bf d}_H(\tau) e^{iq\omega_L \tau}.
\end{equation}

We recall that our analysis has been performed within the single active electron picture. However, in Eqs.~\eqref{eq:def:deltaalpha} and \eqref{eq:def:beta} we have assumed that we have $N$ atoms that contribute to the HHG process coherently in a phase matched way. One can see (see \ref{SM2}) that the shift $\delta \alpha_L(t)$ onto the initial coherent-state is related to the electron and ionization processes taking place in HHG, while the $\beta_q$'s recovers its features regarding the harmonic emission.

\begin{figure*}
    \centering
	\includegraphics[width=1\textwidth]{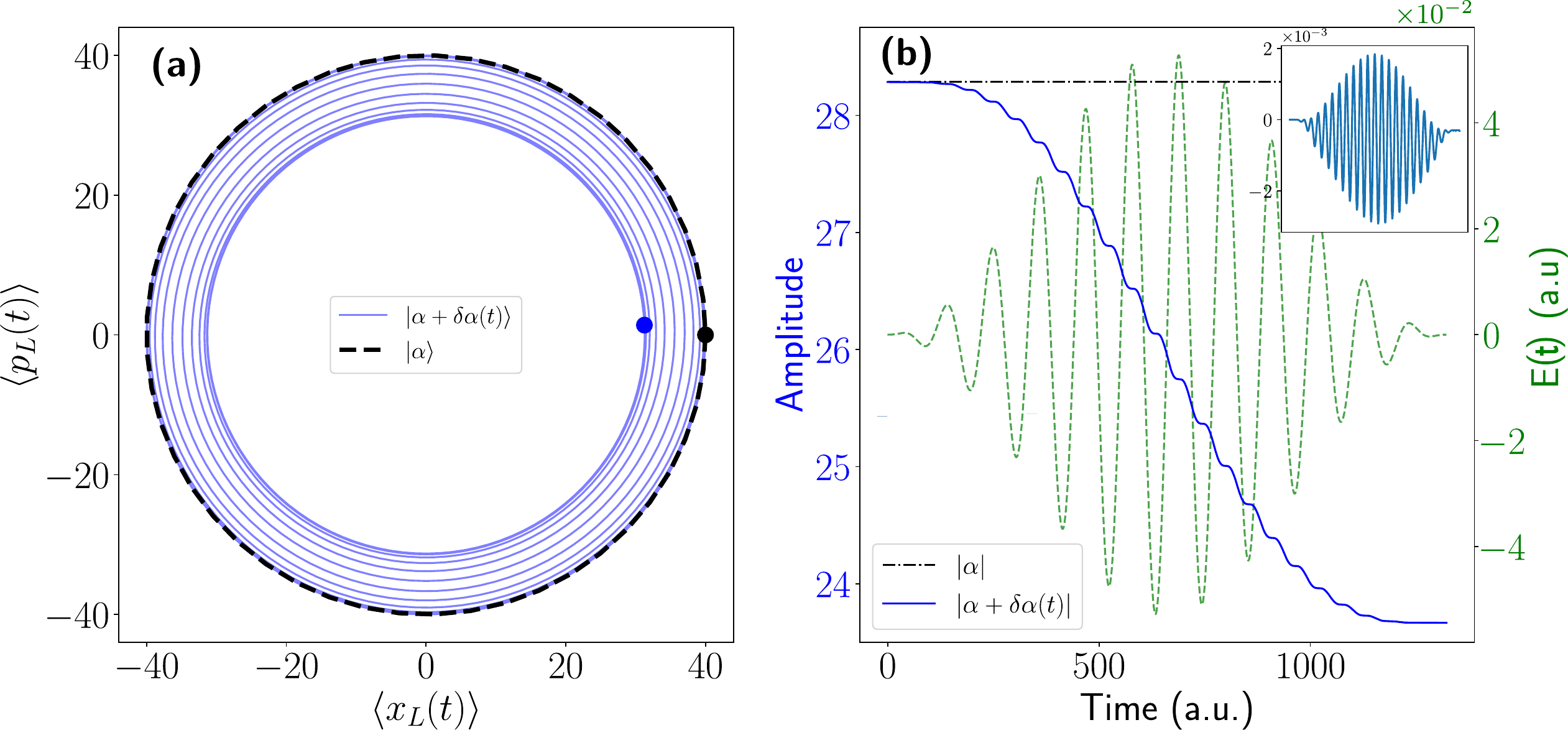}
	\caption{Dynamics of the coherent shifted state obtained after HHG. (a) Dynamics of $\ket{\alpha_L}$ with $\lvert\alpha_L\rvert \approx 28$ (black dashed curve) and $\ket{\alpha_L + \delta\alpha_L(t)}$ (blue continuous curve) in phase space. The analysis was performed using ${\bf g}(\omega_L) \approx 10^{-1}$. The circles in the black dashed and blue continuous curves, which represent $\ket{\alpha_L}$ and $\ket{\alpha_L + \delta \alpha_L(t)}$ respectively, depict the final coherent-state amplitude obtained after the evolution. (b) Dependence of $\lvert\alpha_L + \delta\alpha_L(t)\rvert$ (blue continuous curve) on time. The black dashed line depicts the initial value of $\lvert\alpha_L\rvert$. The applied electric field is plotted with the dashed green line in atomic units (a.u.). The inset plot represents the dependence of the phase $\theta$ with time. We note that in this figure the values of $\alpha_L$ have been chosen in such a way that the effects of $\delta \alpha_L$ could be distinguished. In general, HHG processes take place with values of $\lvert\alpha_L\rvert \approx 10^6$.}  
	\label{Fig2}
\end{figure*}

To study the back-action of the electron acceleration over the initial state of the system, we investigate the phase space dynamics of $\delta \alpha_L$ using the mean value of the photonic quadratures $\hat{x}_L$ and $\hat{p}_L$. Furthermore, we consider the interaction of the laser pulse with a single atom, so that $\delta \alpha_L$ is determined by Eq.~\eqref{eq:def:deltaalpha} when $N = 1$. Defining $\hat{x}_L$, $\hat{p}_L$ as
\begin{equation}\label{Quadratures:def}
    \hat{x}_L
        = \dfrac{1}{\sqrt{2}}
            \big(
                 \hat{a} + \hat{a}^\dagger
            \big)
    \quad \text{and} \quad
    \hat{p}_L
        = \dfrac{1}{i\sqrt{2}}
            \big(
                 \hat{a} - \hat{a}^\dagger
            \big),
\end{equation}
it can be shown that their mean values with respect to Eq.~\eqref{eq:HHG:qstate} are
\begin{equation}\label{quadratures}
    \begin{aligned}
    \langle \hat{x}_L(t) \rangle
        &= \sqrt{2}
            \lvert\alpha_L + \delta \alpha_L(t)\rvert 
            \cos(\omega_L t + \theta (t))\\
    \langle \hat{p}_L(t) \rangle
        &= -\sqrt{2}
            \lvert\alpha_L + \delta \alpha_L(t)\rvert 
            \sin(\omega_L t + \theta (t)),
    \end{aligned}
\end{equation}
with $\theta(t)$ the phase factor of $(\alpha_L + \delta \alpha_L(t)) = \lvert \alpha_L + \delta\alpha_L\rvert e^{-i\theta(t)}$. The integral defining $\delta \alpha_L(t)$ was calculated numerically employing ${\bf d}_H(t)$ extracted from the \textsc{Qprop} software \cite{Qprop}, using a sinusoidal squared laser pulse envelope with 12 cycles and fundamental wavelength $\lambda_L = 800$ nm. Fig.~\ref{Fig2}a shows the amplitude shift of the coherent-state in phase space, while Fig.~\ref{Fig2}b shows the time dependence of the amplitude $|\alpha_L + \delta\alpha_L|$ and the phase factor $\theta(t)$ (inset in Fig.~\ref{Fig2}b). 

The dynamics of $\delta \alpha_L$ is summarized in the following four main features: i) during the acceleration process the ionized electron absorbs photons resulting in an enhancement of $|\delta\alpha_L|$, i.e., a reduction of $|\alpha_L + \delta\alpha_L|$ (see Fig.~\ref{Fig2}b and for more information see \ref{SM3}); ii) $|\delta\alpha_L|$ increases with the amplitude of the driving field as the electron gains more kinetic energy; iii) $|\delta\alpha_L|$ continuously increases during the laser pulse (having a maximum enhancement rate at the peak of the pulse envelope, where the field amplitude is maximum), reaching its maximum value at the end of the pulse; iv) the $|\delta\alpha_L|$ enhancement rate follows the gradient of the driving electric field amplitude. This leads to an oscillatory modulation of frequency $2 \omega_L$ of the enhancement of $|\delta\alpha_L|$ during the laser pulse. It is noted that an oscillatory modulation of frequency $2 \omega_L$ has been also observed on the phase $\theta(t)$ (see inset of Fig.~\ref{Fig2}b). However, because this phase shift is in the order of $10^{-3}$ rad, its influence on the state of the field is considered negligible and, thus, it is not further discussed here. 

Finally, we discuss how the aforementioned results can be used for the creation of optical Schrödinger cat states in the IR spectral region. We note that, although the above analysis is applicable for high-photon-numbers, in the following we discuss the case of low-photon numbers states. This is because we are interested in providing results that can be used by an experiment that utilizes the quantum tomography (QT) method \cite{QT1,QT2} for the quantum state characterization.

To create the coherent-state superposition between the initial coherent-state of the field and its amplitude shifted version, we condition the state of the fundamental field such that it corresponds to the one obtained after HHG as described in ref.~ \cite{Lewenstein1} (see also \ref{SM4}). After reducing the amplitude of the fundamental laser mode, the key action for creating the non-classical light state is the post-selection of the coherent shifted state over those interaction events that lead to the generation of at least one harmonic photon. This is done by performing an anticorrelated measurement between the signal obtained from the harmonic emission and the depletion obtained in the fundamental mode \cite{Tsatrafyllis1,Lewenstein1}. This operation, which we refer to as \emph{conditioning on HHG}, is mathematically expressed for high values of the harmonic cutoff via the projector operator \cite{Stammer2021}
\begin{equation}
    P = \mathbbm{1} - \dyad{\alpha_L}.
\end{equation}

When this operator acts over Eq.~\eqref{eq:HHG:qstate}, and after conditioning the harmonics to be found in $\bigotimes_{q=2}^{\text{cutoff}} \ket{\beta_q}$, the final state of the system is given (up to normalization) by
\begin{equation}\label{eq:HHG:catstate}
    \ket{\Phi_\text{HHG}} = 
        \ket{\alpha_L + \delta \alpha_L}
        - \xi \ket{\alpha_L},
\end{equation}
which is the superposition of two coherent-states, commonly referred to as optical cat states \cite{Brune1992, Deleglise2008}, where $\xi \approx \braket{\alpha_L}{\alpha_L + \delta \alpha_L}$. Note that the dependence of the weight $\xi$ with $\delta \alpha_L$ allows us to control the quantum features of the state \cite{Rivera2021}, for example by modifying the density of atoms in the interaction region or the intensity of the employed laser field. In particular, in the limit where $\delta \alpha_L \to 0$ we get an optical ``kitten'' state characterized by $\ket{\Phi_\text{HHG}} \approx D(\alpha_L) \ket{1}$ (see \ref{SM4}), while the limit $0<\xi<1$ leads us to the ``genuine cat'' state presented in Eq.~\eqref{eq:HHG:catstate}. Furthermore, if $\lvert\delta \alpha_L\rvert$ becomes a very large quantity such that $\xi \to 0$, then the final state is just given by the amplitude shifted coherent-state $\ket{\Phi_\text{HHG}} = \ket{\alpha_L + \delta \alpha_L}$.

\begin{figure}
	\includegraphics[width=1.\columnwidth]{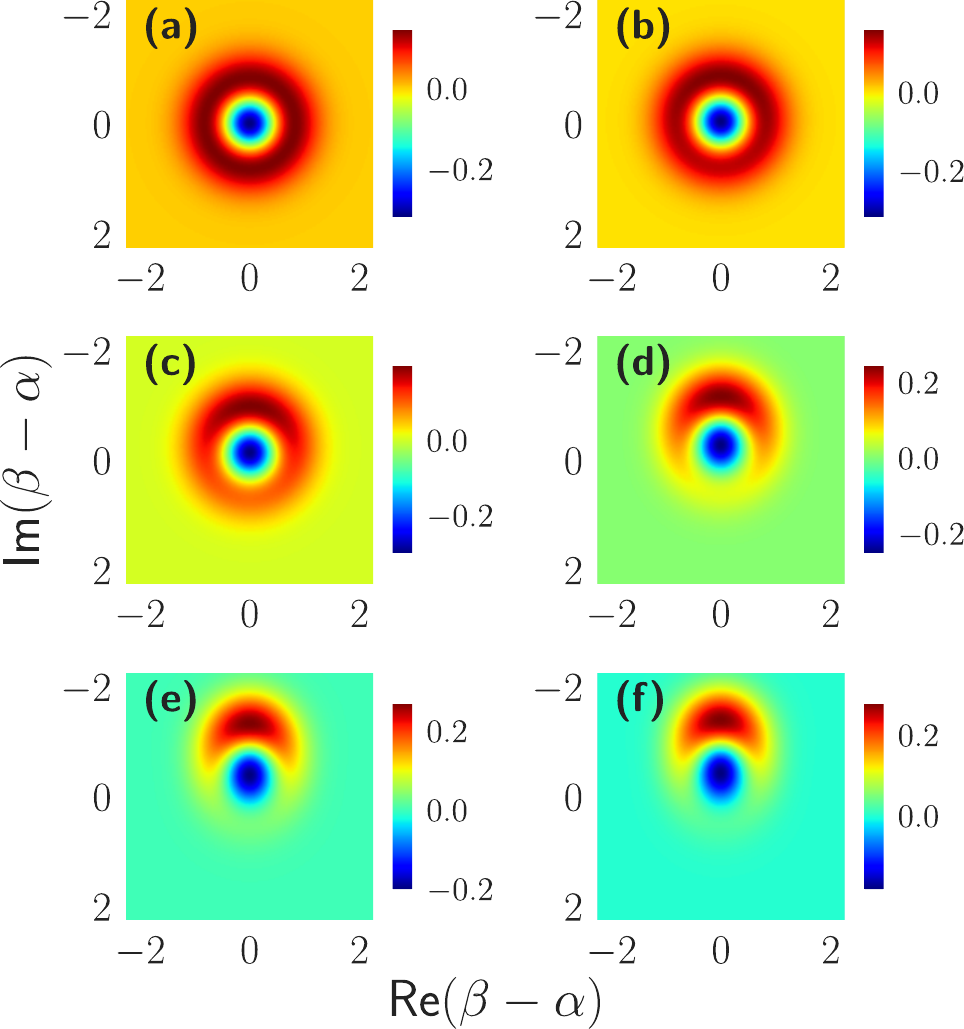}
	\caption{Wigner function evolution along the pulse when conditioning to HHG. Here, we have used the same electric field as in Fig.~\ref{Fig2}, and see how the Wigner function looked after (a) 2, (b) 4, (c) 6, (d) 8, (e) 10 and (f) 12 cycles. $\Re[\beta-\alpha]\equiv x_L, \Im[\beta-\alpha]\equiv p_L$, with $x_L$, $p_L$ the values of the quadrature field operators $\hat{x}_L=(\hat{a}+\hat{a}^\dagger)/\sqrt{2}$ and $\hat{p}_L=(\hat{a}-\hat{a}^\dagger)/i\sqrt{2}$.}
	\label{Fig3}
\end{figure}

The different cases discussed above are shown in Fig.~\ref{Fig3}, where the Wigner function of the final state (see \ref{SM4}) has been calculated using an electric field with a sinusoidal squared envelope, 12 cycles of duration ($\sim 30$ fs duration), wavelength $\lambda_L = 800$ nm and amplitude $E_L= 0.053$ a.u., where a.u. denotes atomic units (which corresponds to a laser intensity of about $10^{14}$ Watt per cm$^2$). At the beginning of the pulse (Fig.~\ref{Fig3}a--b), where the driving field amplitude is small, $\delta \alpha_L$ is small resulting to the creation of a ``kitten'' state, while at the end of the pulse where $\delta \alpha_L$ is getting larger (according to Fig.~\ref{Fig2}b) the Wigner function depicts a genuine ``cat'' state (Fig.~\ref{Fig3}e--f). Evidently, in case of reducing the intensity of the driving field the final state would be a ``kitten''.

One of the main advantages of using the Wigner function as an observable for the final quantum optical state of the field is that it allows the superposition between the two coherent-states to be seen explicitly. If the depletion in the fundamental is small enough to witness a Wigner negativity, the anticorrelation measurement is not able to exactly distinguish the contributions from the depleted field $\ket{\alpha_L + \delta \alpha_L}$, and the input state $\ket{\alpha_L}$. It is the indistinguishability at the detector which leads to the observed interference. In particular, if $\delta \alpha_L$ is too small, we get the ``kitten'' case where both states contribute equally to the Wigner function, and the distribution has an homogeneous ring-like shape structure around a negativity center that witnesses the quantum superposition. As the depletion increases, we get a genuine ``cat'' for which the distribution is not homogeneous, since the contribution of $\ket{\alpha_L + \delta \alpha_L}$ is bigger than the one provided by $\ket{\alpha_L}$ because of the $\xi$ prefactor. This increasing in the distinguishability leads to smaller values of the Wigner negativity. Finally, in the case of enormous values of the depletion (leading to $\xi \to 0$), we only observe the contribution from $\ket{\alpha_L + \delta \alpha_L}$ and, hence, the Wigner function is a Gaussian centered around $\alpha_L + \delta \alpha_L$.

\subsection{Quantum optical dynamics of ATI}\label{Section:ATI}
As mentioned above, ATI processes occur when the ionized electron either does not re-collide with the parent ion or, if it does, the process takes place elastically. Therefore, to study these phenomena within our formalism, we will condition Eq.~\eqref{eq:total:TDSE} upon finding the electron in continuum states, which we will simply represent as $\ket{\bf v}$, where ${\bf v}$ denotes the outgoing kinetic momenta of the electron. In this case, the conditioned Schrödinger equation reads
\begin{equation}\label{eq:ATI:TDSE}
    \begin{aligned}
	    i \hbar \pdv{}{t} \braket{\bf v}{\psi(t)}
		    = -\opE_Q(t) \cdot \mel{\bf v}{\text{e}\opr_H(t)}{\psi(t)}.
	\end{aligned}
\end{equation}

At this point, we introduce the SFA theory assumptions \cite{Lewenstein1994} and neglect the effects of the electronic bound excited states. Thus, introducing the SFA version of the identity
\begin{equation}\label{eq:SFA:Id}
    \mathbbm{1}
		\approx \dyad{\text{g}} + \int \dd {\bf v} \dyad{\bf v},
\end{equation}
in Eqs.~\eqref{eq:HHG:TDSE} and \eqref{eq:ATI:TDSE}, we get the following set of coupled differential equations
\begin{equation}\label{Coupled:ATI}
	\begin{aligned}
	&i \hbar \pdv{}{t} \ket{\Phi(t)}
		= -\opE_Q(t)\cdot\meld_H(t) \ket{\Phi(t)}\\
			&\hspace{2.1cm}
			- \int\! \dd {\bf v}\ \opE_Q(t)\cdot \meld_H({\bf v},t) \ket{\Phi({\bf v},t)}\\
	& i \hbar\pdv{}{t}\ket{\Phi({\bf v},t)}
		= -\opE_Q(t)\cdot\meld^*_H({\bf v},t) \ket{\Phi(t)}\\
		    &\hspace{2.45cm}
		    - \int\! \dd {\bf v'} \opE_Q(t)\cdot \meld_H({\bf v},{\bf v'},t) \ket{\Phi({\bf v'},t)},
	\end{aligned}
\end{equation}
where we denote the conditioned to ATI state as $\ket{\Phi({\bf v},t)} = \braket{{\bf v}}{\psi(t)}$, $\meld_H({\bf v},t) = \mel{\bf v}{\text{e}\opr_H(t)}{\text{g}}$ the time-dependent dipole moment matrix element between states $\ket{\bf v}$ and $\ket{\text{g}}$, and $\meld_H({\bf v},{\bf v'},t) = \mel{\bf v}{\text{e}\opr_H(t)}{\bf v'}$ represents the time-dependent dipole moment matrix element between states $\ket{\bf v}$ and $\ket{\bf v'}$.

In the spirit of the SFA theory, we may neglect the effect of the continuum-continuum transitions and obtain the contribution to ATI corresponding to direct tunnelling, or either treat the continuum-continuum transitions perturbatively \cite{Kulander,Noslen1,Noslen2,Noslen3} in order to describe the rescattered ATI electrons at higher energies up to $10U_p$, where $U_p$ is the ponderomotive potential defined as $U_p = \text{e}^2E^2/4m\omega_L^2$ with $E$ the electric field amplitude and $m$ the electron's mass. Thus, considering electrons of ``low'' kinetic energy ($<2U_p$) and keeping the strong-field approximations, the state conditioned to ATI reads
\begin{equation}\label{eq:ATI:single}
	\ket{\Phi({\bf v},t)}
		= i\hbar \int_{t_0}^t \dd t' \
			\opE_Q(t') \cdot \meld^*_H({\bf v},t') \ket{\Phi(t')},
\end{equation}
where $\ket{\Phi(t)}$ is the solution to Eq.~\eqref{eq:HHG:TDSE}.

To derive the reduced density matrix for the electromagnetic field that corresponds to ATI processes, we will consider two different strategies: (i) we condition on ATI electrons that have a specific outgoing direction and kinetic momentum $\vb{v}$, which leads to a reduced density matrix of the form $\rho = \dyad{\Phi({\bf v},t)}$ (pure state); (ii) we condition on all possible ATI electrons without distinguishing on the particular direction and kinetic momentum of the outgoing electrons, which leads to $\rho = \int \dd^3 v \dyad{\Phi({\bf v},t)}$ (mixed state). 

\begin{figure}
	\includegraphics[width= \columnwidth]{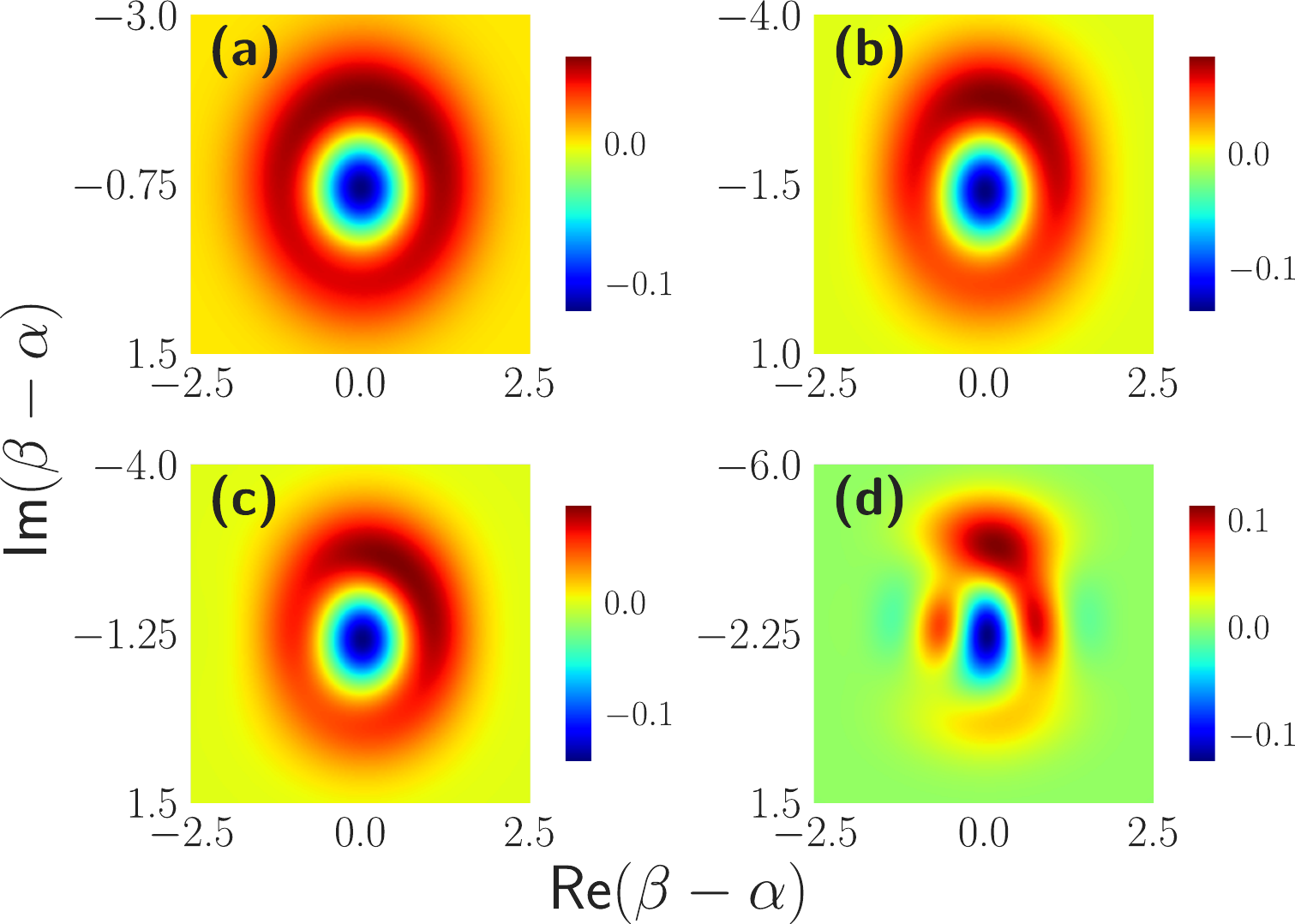}
	\caption{Wigner function after single-ionization ATI. Dependence of the Wigner function with the number of half-cycles $\mathcal{N}$ of equal intensity provided by a laser source, and with the shift between two consecutive coherent-states $\Delta$. In these subplots we consider: (a) $\mathcal{N} = 5, \Delta = -0.25i$; (b) $\mathcal{N} = 5, \Delta = -0.5i$; (c) $\mathcal{N} = 8, \Delta = -0.25i$; (d) $\mathcal{N} = 8, \Delta = -0.5i$. $\Re[\beta-\alpha]\equiv x_L, \Im[\beta-\alpha]\equiv p_L$, with $x_L$, $p_L$ the values of the quadrature field operators $\hat{x}_L=(\hat{a}+\hat{a}^\dagger)/\sqrt{2}$ and $\hat{p}_L=(\hat{a}-\hat{a}^\dagger)/i\sqrt{2}$.}
	\label{Fig4}
\end{figure}

For the first scenario, assuming that during the ATI process the harmonic coherent-state amplitudes ($\beta_q$ in Eq.~\eqref{eq:HHG:qstate}) stay very close to the vacuum, one can see that the final state of the system can be written as (for more details see \ref{SM5})
\begin{equation}\label{eq:ATI:SingleState}
    \begin{aligned}
    \ket{\Tilde{\Phi}(\textbf{v},t)}
        &\approx i\hbar \sum_{j=0}^{\mathcal{N}-1}
        \int^{t_{j+1}}_{t_j} \! \! \! \! \! \dd t'\
        \opE_L(t') \cdot \meld^*_H({\bf v},t')\\
        & \hspace{4.1cm}
            \times \ket{(j+1)\Delta},
    \end{aligned}
\end{equation}
where $\mathcal{N}$ is the number of half-cycles and $\Delta$ is the amount of photons absorbed in each half-cycle (as discussed in Fig.~\ref{Fig2}). We see that the final state is given as a superposition of different coherent-states (which in principle is larger than two), where each of them is affected by the instantaneous value of the electric field operator evaluated at time $t'$. In Fig.~\ref{Fig4}, we present the Wigner functions calculated from Eq.~\eqref{eq:ATI:SingleState}. In these calculations, we assumed that the electron tunnels out with zero kinetic energy, and considered (a) $\mathcal{N} = 5, \Delta = -0.25i$, (b) $\mathcal{N} = 5, \Delta = -0.5i$, (c) $\mathcal{N} = 8, \Delta = -0.25i$ and (d) $\mathcal{N} = 8, \Delta = -0.5i$. As we can see, as both $\mathcal{N}$ and $\Delta$ increase, the distance between the two outermost coherent-states appearing in the superposition also increases and we switch from a kitten state (like the one in Fig.~\ref{Fig4} (a)) to more complicated coherent-state superpositions (like the one in Fig.~\ref{Fig4} (d)). Note that the distribution shown in Fig.~\ref{Fig4} (d) differs from the symmetric one  coming from a coherent-state superposition of the form $\ket{\alpha}\pm \ket{-\alpha}$, in that we have more states in the superposition which are contributing as well to the Wigner function.

Another difference that we observe in these plots is that some of the Wigner distributions obtained for single-ionization ATI depict a small rotation (see for instance Fig.~4c). This is related to a change in the phase of the coherent-states appearing in the superposition. However, it may also be the case that small rotations are related to a change in the phase of the respective amplitudes in the superposition, which at the end is related on how we are implementing the conditioning operations. In HHG, the coefficient $\xi$ appears as a consequence of the conditioning measurement that is being applied to the optical modes, and if either both $\delta \alpha$ and $\alpha$ have the same phase or a phase difference of $\pi$, as it happens in the present manuscript, then $\xi$ is a real quantity. In ATI, the coefficients weighting the superposition have a different nature, as they depend, via $\vb{d}^*(\vb{v},t)$, on the electron's trajectory before being detected, which in general is a complex quantity. Thus, in single-ionization ATI we might find changes in the coefficients from one term to the other, leading to these rotations. Related to this, we expect that one of the main effects of the carrier-envelope phase, i.e. the change of phase between the carrier wave and the field envelope, over the final Wigner distribution is the presence of these rotations, which would affect the HHG state. However, further research has to be done in this direction, since our analysis is restricted to a multicycle pulse.

\begin{figure}
	\includegraphics[width= \columnwidth]{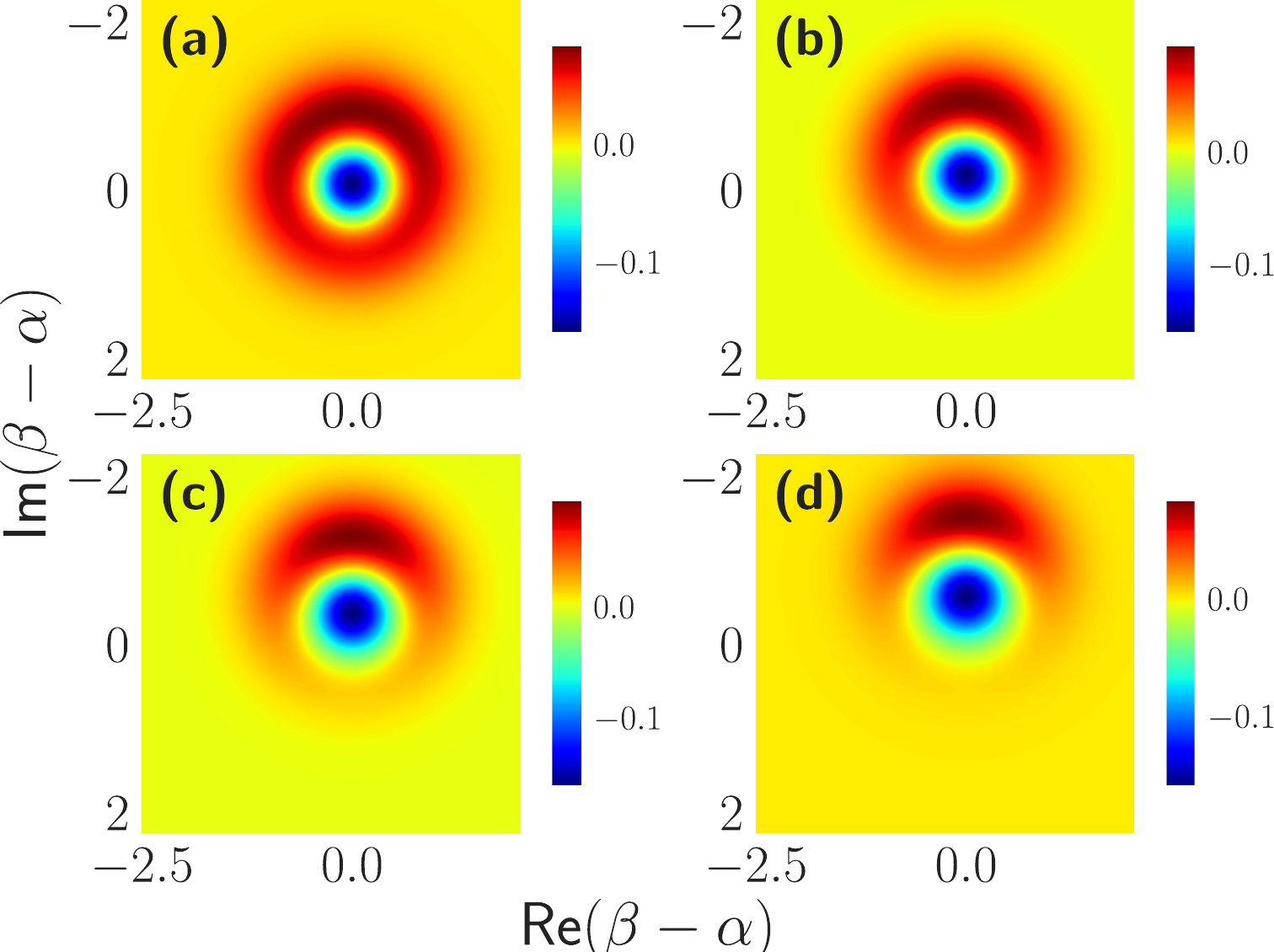}
	\caption{Wigner function after ATI and conditioning over all possible momenta. Calculated Wigner functions after considering equal and time-independent coherent-shifts (a) $\delta \alpha = -0.1i$, (b) $\delta \alpha = -0.25i$, (c) $\delta \alpha = -0.5i$ and (d) $\delta \alpha = -0.75i$. For the computation of the Wigner function we have further considered some approximations over the time-dependent integrals, which are detailed in \ref{SM5}. $\Re[\beta-\alpha]\equiv x_L, \Im[\beta-\alpha]\equiv p_L$, with $x_L$, $p_L$ the values of the quadrature field operators $\hat{x}_L=(\hat{a}+\hat{a}^\dagger)/\sqrt{2}$ and $\hat{p}_L=(\hat{a}-\hat{a}^\dagger)/i\sqrt{2}$.} 
	\label{Fig5}
\end{figure}

For the second scenario, in order to gain intuition about the obtained mixed state, we are going to consider a linearly polarized field, and assume: (i) that during the ATI process the harmonic coherent-state amplitudes stay very close to the vacuum, and (ii) that the generated coherent shifts are identical and time-independent. In general this is not true and, as discussed in Fig.~\ref{Fig2}b, the IR coherent-state is continuously increasing (in modulus) along the pulse. However, for single-electron ionization processes, one may expect this shift to remain very small. Therefore, under these considerations the ATI state conditioned to all outgoing momenta reads
\begin{equation}\label{eq:ATI:MixedState}
    \begin{aligned}
    \rho_{\text{ATI-IR}}
        &= \int^t_{t_0}\dd t' \int^t_{t_0}\dd t''
            \hat{E}_L(t') \dyad{\delta \alpha} \hat{E}_L(t'')\\
                & \hspace{2.5cm}
                \times K(t',t'') e^{i\varphi(t')}e^{-i\varphi(t'')},
    \end{aligned}
\end{equation}
where $\hat{E}_L$ is the part of the electric field operator in Eq.~\eqref{eq:Efield} that acts over the fundamental mode and $K(t,t'') = \langle \hat{d}_H(t')\hat{d}_H(t'')\rangle - \langle \hat{d}_H(t')\rangle \langle \hat{d}_H(t'')\rangle$ (see \ref{SM5}). The results for the calculated Wigner functions are shown in Fig.~\ref{Fig5}, where in each of the subplots we have considered increasing values of $\delta \alpha$ (from (a) to (d)). We note that its shape is very similar to a ``cat'' state and, as it happens in HHG, as $\delta \alpha$ increases it tends to a typical Gaussian state. This is due to the approximations we considered and that lead to Eq.~\eqref{eq:ATI:MixedState}, since in the limit when $\delta\alpha$ is very big we can write $\hat{E}_L(t) \ket{\delta \alpha} \propto \ket{\delta\alpha}$ which leads to the Gaussian-like Wigner function. However, we note that this limit is not compatible with our assumptions, since we expect $\delta \alpha$ to be small in the single active electron picture. More non-classical features are expected for the exact state obtained after the interaction, i.e. without approximations, due to the change of $\delta \alpha$ in time. We also note that the rotations obtained in the Wigner distributions appearing in \emph{single-ionization ATI} do not show up in this case. Although this is an expected feature given that the $K(t,t')$ is a complex function, the approximations we consider here in order to gain intuition about the shape of the final Wigner functions, do not take account for it.

Finally, we remark that the plots we have presented thus far for the ATI process correspond to the state right after the interaction, i.e., in the displaced frame of reference. However, for the Wigner function characterization this is not a problem as, by implementing them, one observes the same features as the ones shown in our figures upon a shift and a rotation.

\section{EXPERIMENTAL RESULTS}
\subsection{Experimental setup}\label{Sect:ExperimentalSetup}

The quantum features of the non-classical light state of the fundamental mode exiting the atomic medium depends on the used conditioning approaches (HHG and/or ATI) and on $\delta\alpha_L$, which introduces the dependence with the gas pressure in the interaction area (Eq.~\eqref{eq:def:deltaalpha}). Here, the action of conditioning was achieved using the quantum spectrometer (QS) approach \cite{Tsatrafyllis1, Tsatrafyllis2} and the quantum state characterization was performed by means of homodyne detection and the well known QT method \cite{QT1, QT2}. In the following,  after the description of the operation principle of the experimental approach (see also ref. \cite{Lewenstein1} and \ref{SM6}), we experimentally demonstrate the dependence of the coherent-state superposition (created by conditioning on HHG) on $\delta\alpha_L$, and the generation of high-photon-number optical ``cat'' states. Following a similar strategy, the method can be used for the characterization of optical coherent-state superpositions generated by conditioning on the ATI process (see \ref{SM5} and \ref{SM7}). This can be achieved by using the ATI photoelectron signal recorded by means of a time-of-flight electron spectrometer (see \ref{SM7}, Fig.~\ref{FigSM2}).

\begin{figure}
	\includegraphics[width= \columnwidth]{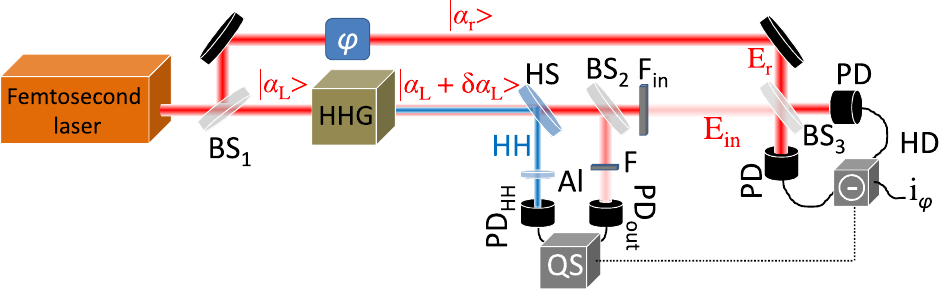}
	\caption{Simplified scheme of the experimental set-up. $\ket{\alpha_L}$ and $\ket{\alpha_r}$ are the IR coherent-states transmitted and reflected by an IR beam separator BS$_{1}$. The transmitted IR beam is focused into a xenon gas jet where the high--harmonics (HH) are generated. $\ket{\alpha_L+\delta\alpha_L}$ is the state of the IR field after the interaction. HS is a harmonic separator which reflects the HH and leaves the IR beam to pass through. BS$_{2,3}$ is an IR beam separator and splitter, respectively.  PD$_{out}$ and PD$_{HH}$ are the IR and HH photodetectors, respectively, used by the QS to condition the IR field exiting the atomic medium on the HHG. Just before PD$_{HH}$ a 150 nm thick aluminum filter was placed in order to select the harmonics with $q \geq 11$ and block any residual part of the IR beam. $F$ and $F_{in}$ are neutral density filters. $E_{in}$ is the state of the IR field to be characterized. PD are the IR photodetectors used by the balanced detector of the homodyne detection (HD) system. $E_{r}$ is the field of the reference beam. $\varphi$ is the controllable phase shift introduced in the reference beam and $i_{\varphi}$ is the photocurrent difference which is proportional to the measurement of $\hat{x}_{\varphi}$. When the xenon gas jet and the QS was switched on the homodyne detection system provides the measurement $\hat{x}_{\varphi}$ only when IR field exiting the atomic medium is conditioned on, the HHG and via QT provides the Wigner function of the light state $\ket{\Phi_\text{post}}=\ket{\alpha_L+\delta\alpha_L}- \xi \ket{\alpha_L}$ with $\xi=\braket{\alpha_L}{\alpha_L +\delta\alpha_L}$. }
	\label{Fig6}
\end{figure}

A schematic of the experimental approach is shown in Fig.~\ref{Fig6}. The experiment was performed using as a primary laser source a Ti:Sapphire laser system delivering linearly polarized $\approx$ 35 fs pulses of $\lambda\approx$ 800 nm carrier wavelength. The IR laser beam was separated into the branches of an interferometer by a beam separator BS$_1$. The reflected IR beam serves as a reference beam of the QT method. The transmitted IR beam was focused with an intensity $\approx 8 \times 10^{13}$ W/cm$^2$ into a xenon pulsed gas jet, where harmonics up to 21st order have been generated. The photon number of the generated XUV beam (reflected by a harmonic separator HS) and the photon number of a portion the IR beam (reflected by an IR beam separator BS$_2$), have been recorded for each laser shot by the PD$_{HH}$ and PD$_{out}$ photo detectors, respectively. These were used by the QS to condition the IR field exiting the atomic medium on the HHG process (see \ref{SM6}). After BS$_2$, the mean photon number of the IR field was reduced (by means of neutral density filters F$_{in}$) to the level of few photons per pulse. The IR field amplitude before reaching the balanced detector of the homodyne detection system is denoted with E$_{in}$. The E$_{in}$ field was spatiotemporally overlapped in a beam splitter (BS$_3$) with the high-photon-number reference field E$_{r}$ coming from the second branch of the interferometer. The interfering fields after BS$_3$ were recorded by a balanced detector, which provides at each value of $\varphi$ for each laser shot the photocurrent difference $i_{\varphi}$. The values of $i_{\varphi}$ are directly proportional to the measurement of the electric field operator $\hat{E}_{in} (\varphi) \propto \hat{x}_{\varphi}=\cos(\varphi) \hat{x}+ \sin(\varphi) \hat{p}$, and have been used for the reconstruction of the Wigner function refs. \cite{QT2,Leonhardt,Breitenbach1997} (see \ref{SM7} and \ref{SM8}). When the xenon gas jet and the QS were switched on, the homodyne detection system measures the $\hat{x}_{\varphi}$ only when the IR field exiting the atomic medium is conditioned on the HHG, providing via QT the Wigner function of the light state $\ket{\Phi_\text{post}}=\ket{\alpha_L+\delta\alpha_L}- \xi \ket{\alpha_L}$ with $\xi=\braket{\alpha_L}{\alpha_L +\delta\alpha_L}$. 

\begin{figure}
	\includegraphics[width= \columnwidth]{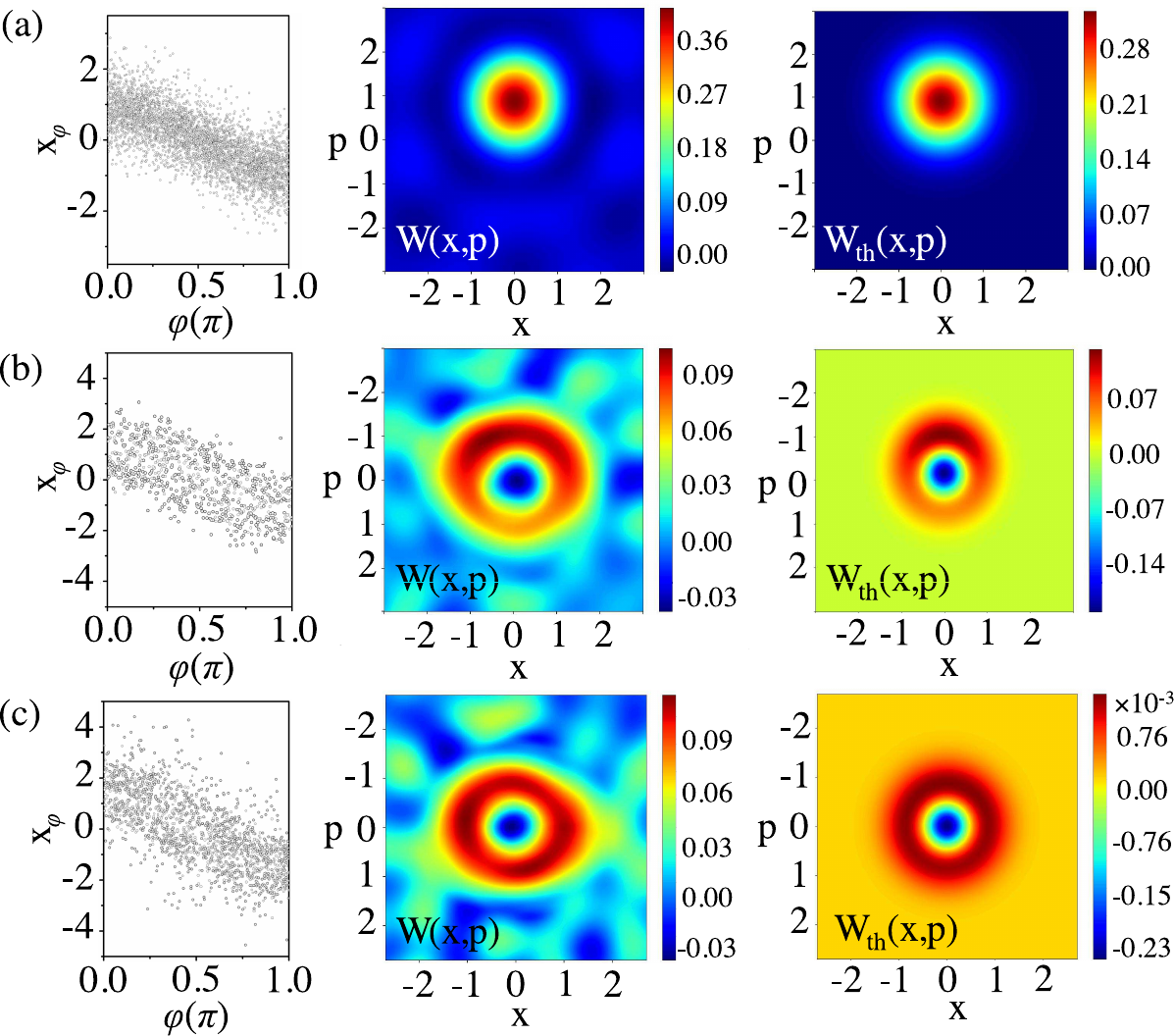}
	\caption{Optical ``cat'' and ``kitten'' states created by conditioning on HHG, for different values of $\lvert\delta \alpha_L\rvert$. The left, middle and right panels show the measured $\hat{x}_{\varphi}$, the corresponding reconstructed $W(x,p)$, and the theoretically calculated $W_{th}(x,p)$, respectively, projected onto the $(x, p)$ plane. (a) coherent-state of driving laser field measured when the Xe gas and QS approach were switched off. (b) Optical ``cat'' state measured when the Xe gas jet and the QS were switched on and the harmonic yield was close to maximum. The corresponding $W_{th}(x,p)$ has been calculated for $|\delta \alpha_L|\approx 0.5$, where $|\alpha_L|\approx1.4$ and $|\xi|\approx0.88$. (c) Optical ``kitten'' state measured when the the harmonic yield was reduced by a factor of $\approx25$, i.e., $\delta\alpha_L$ by a factor of $\approx5$, compared to the harmonic yield of (b). The corresponding $W_{th}(x,p)$ has been calculated for $|\delta \alpha_L|\approx0.1$, where $|\alpha_L|\approx1.3$ and $|\xi|\approx0.99$. $x$ and $p$ are the values of the quadrature field operators $\hat{x}=(\hat{a}+\hat{a}^\dagger)/\sqrt{2}$ and $\hat{p}=(\hat{a}-\hat{a}^\dagger)/i\sqrt{2}$. The Wigner functions in these plots have been centered around the value of $\alpha_L$.}
	\label{Fig7}
\end{figure}

\begin{figure*}
	\includegraphics[width= 1\textwidth]{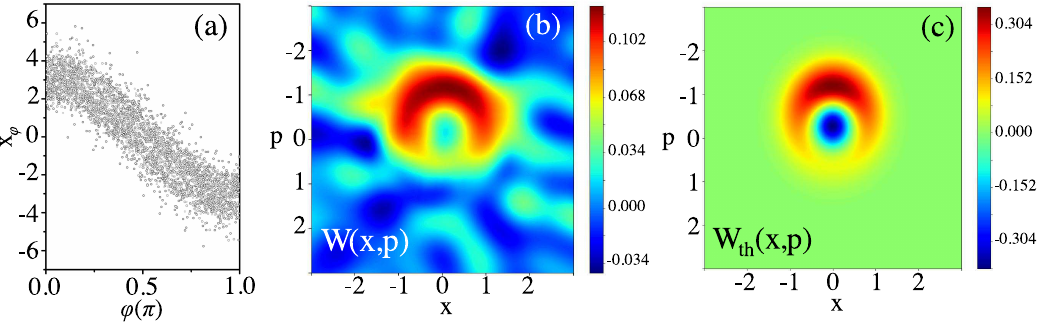}
	\caption{high-photon-number optical ``cat'' state created by conditioning on HHG. (a) Measured $\hat{x}_{\varphi}$ with xenon gas and QS switched on. (b) Projection on $(x, p)$ plane of the reconstructed $W(x,p)$ which shows an optical ``cat'' state of $\langle{n}\rangle\approx9.4\pm0.1$. (c) Theoretically calculated Wigner function $W_{th}(x,p)$ for $|\delta \alpha_L|\approx0.8$, where $|\alpha_L|\approx3.7$ and $|\xi|\approx0.73$. $x$ and $p$ are the values of the quadrature field operators $\hat{x}=(\hat{a}+\hat{a}^\dagger)/\sqrt{2}$ and $\hat{p}=(\hat{a}-\hat{a}^\dagger)/i\sqrt{2}$. The Wigner functions in these plots have been centered around the value of $\alpha_L$.}
	\label{Fig8}
\end{figure*}
\subsection{Dependence of the coherent-state superposition on $\delta\alpha_L$: Optical ``kitten'' and ``cat'' states}\label{Section:density}
To show the dependence of the quantum features of the coherent-state superposition with $\delta\alpha_L$, we have measured the Wigner function $W(x,p)$ for two different values of $\delta\alpha_L$ when we condition on HHG. This is shown in Fig.~\ref{Fig7} together with the measurement of the coherent-state of the driving field (Fig.~\ref{Fig7}a). The left panels show the measured $\hat{x}_{\varphi}$, the middle panels the corresponding reconstructed $W(x,p)$, and the right panels the theoretically calculated $W_{th}(x,p)$. As $\delta\alpha_L \propto N$ (Eq.~\eqref{eq:def:deltaalpha}), the change of $\delta\alpha_L$ was achieved by varying the number of atoms $N$ in the interaction region (using the delay between the laser pulse arrival and the opening of the Xe gas nozzle). It is noted that for experimental reasons (gas load in the vacuum chamber), in the present experiment the maximum value of the used $N$ was set such that the harmonic signal was slightly lower (a factor of $\approx 2$) than its maximum value. Since the harmonic yield ($Y$) is $Y\propto N^2$, we then get $\delta\alpha_L\propto Y{^{1/2}}$. This relation provides a useful experimental guide for controlling the value of $\delta\alpha_L$ by monitoring the integrated signal of the harmonics passing through the Aluminum filter.

For reasons of completeness and for evaluating the performance of the experimental setup, it is useful to measure first the coherent-state of the driving field by switching off the Xe gas jet and the QS. This is shown in Fig.~\ref{Fig7}a. As expected, the state of the IR driving field is coherent, depicting a $W(x,p)$ with Gaussian distribution. The same result was obtained when the Xe gas and the QS were switched on and off, respectively. By switching on both, the Xe gas jet (at conditions where the harmonic generation yield is close to maximum) and the QS, as reported in ref  \cite{Lewenstein1}, an optical ``cat'' state with mean photon number $\langle{n}\rangle\approx1.74\pm0.03$ has been recorded (Fig.~\ref{Fig7}b). The $W(x,p)$ depicts a half--ring--like shape with a central negative minimum located at $(x_{min}, p_{min}) \approx (0, 0)$ and a maximum at $(x_{max}, p_{max}) \approx (0, -1)$, which is in agreement with the $W_{th}(x,p)$ obtained by the theoretical calculations for $|\delta \alpha_L|$ in the range of $0.4$ to $0.5$. In Fig.~\ref{Fig7}b we show the $W_{th}(x,p)$ for $\lvert\delta \alpha_L\rvert\approx0.5$, where $|\alpha_L|\approx1.4$ and $\lvert\xi\rvert \equiv \lvert\braket{\alpha_L + \delta\alpha_L}{\alpha_L}\rvert\approx0.88$. The value of $|\alpha_L|$ has been obtained by the equation $\langle n \rangle = \mel{\Phi_\text{post}}{\hat{n}}{\Phi_\text{post}}$ using as $\langle n \rangle$ the value of the measured mean photon number. When we reduce the $Y$ by a factor of $\approx25$, i.e., $\delta\alpha_L$ by a factor of $\approx5$, the state superposition transitions from an optical ``cat'' to ``kitten'' state. This is shown in Fig.~\ref{Fig7}c where an optical ``kitten'' state  with $\langle{n}\rangle\approx2.54\pm0.05$ has been recorded. In this case, the measured $W(x,p)$ depicts a full--ring shape with a central negative minimum located at $(x_{min}, p_{min}) \approx (0, 0)$. This is in agreement with the $W_{th}(x,p)$ obtained by the theoretical calculations obtained for $|\delta \alpha_L|\approx 0.1$, where $\lvert\alpha_L\rvert\approx1.3$ and $\lvert\xi\rvert\approx0.99$. We note that, for values of $\lvert\delta  \alpha_L\rvert < 0.1$, our cat state behaves as a displaced Fock state, as there is no pronounced maximum on the ring shape phase space distribution.

\subsection{Generation of high-photon-number optical ``cat'' states}\label{Section:HighCats}
For applications in quantum technology it is also important to be able to increase the photon number of the produced optical ``cat'' states. As was mentioned before, the present approach can be used for the production of arbitrary high-photon-number ``cat'' states. To show this, we have recorded a 9-photon shifted optical ``cat'' state (Fig.~\ref{Fig8}) created by conditioning on HHG. Fig.~\ref{Fig8}a shows the measurement of $\hat{x}_{\varphi}$ used to reconstruct the Wigner function shown in phase space in Fig.~\ref{Fig8}b. The measurement was performed using a value of $N$ approximately close to the value used to record the low-photon number optical ``cat'' state shown in Fig.~\ref{Fig7}b, while the photon number has been increased by means of $F_{in}$ (Fig.~\ref{Fig6}). This was achieved by the fine adjustment of the angle of the $F_{in}$ filter with respect to the incoming beam. In this case, an optical ``cat'' state with $\langle{n}\rangle\approx9.4\pm0.1$ has been recorded. The $W(x,p)$ depicts a half--ring--like shape with a central minimum located at $(x_{min}, p_{min}) \approx (0, -0.2)$ and a maximum at $(x_{max}, p_{max}) \approx (0, -1.2)$. The shape of the measured $W(x,p)$ is reasonably close to the Wigner function ($W_{th}(x,p)$) obtained by the theoretical calculations for $|\delta \alpha_L|$ in the range of $0.6$ to $1.1$. In Fig.~\ref{Fig8}c, we show the $W_{th}(x,p)$ for $|\delta \alpha_L|\approx 0.8$, where $|\alpha_L|\approx3.7$ and $|\xi|\approx0.73$. The lack of negative values at the position of the minimum of the measured $W(x,p)$, is attributed to limitations of the present experimental approach in obtaining the Wigner function and the photon number with accuracy better than $\pm0.004$ and $\approx1.5\%$, respectively (see \ref{SM7} and \ref{SM8}). 

The limitations introduced for further increasing the mean photon number of the shifted optical cat state are associated with the resolution of the detection system and the decoherence effects (see \ref{SM8} and \ref{SM9} respectively), which cannot be excluded. A quantitative analysis of the decoherence effects and their dependence on the photon number of the lightstate, requires an extensive theoretical and experimental investigation which is out of the scope of our work. The present results cannot be used for such analysis. However, in order to further stress the potential of our approach to produce high-photon-number shifted optical cat states in a lossy environment, we have used a simple, although exact, noise model that introduces photon losses due to the interaction with a Gaussian reservoir \cite{Leonhardt1993}. This is done by means of a beam splitter where in one of the inputs we introduce our cat state, while on the other an ancillary vacuum mode that is later on traced out (for more details see \ref{SM9}). This model shows that, even in the case of high photon losses (in the range of 60\%), although the negativity of the Wigner function of the optical cat state is reduced, the main features are maintained.

\section{DISCUSSION AND PERSPECTIVES}\label{Discussion}
In the last two decades, pioneering optical methods in quantum state engineering have been implemented for the generation of optical cat-like and cat states (c.f. \cite{Ourjoumtsev2006,Ourjoumtsev2007,Hacker2019,Zavatta2004,Sychev2017}). These  methods rely on the use of few photon number and high-fidelity Fock states primary sources and currently deliver optical cat states in the range of few photon numbers, restricting their applicability in quantum technologies. This is because the quantum technology toolbox contains passive linear optical elements (such as phase shifters, beam splitters and fiber optics), which unavoidably have optical losses. Thus, it is evident that any beam propagating through these elements will naturally suffer from photon losses. Hence, one of the main motivations for generating high-photon-number optical cat states (as we report here), is associated with their power to be used in more complex optical arrangements that can lead to the generation of large optical cat states and massively entangled state superpositions with controllable quantum features. Such states could highly benefit from investigations concerning the fundamental tests of quantum theory, quantum information processing, metrology/sensing, and communication. Towards these directions, we have recently reported how the method presented here, can be used for the development of more complex optical arrangements that can lead to generation of i) controllable large coherent-state superpositions \cite{Rivera2021}, and ii) multimode entangled states spanning from the near infrared to the extreme ultraviolet \cite{Stammer2021}, which can be very useful for quantum technology.

Additionally, and in a more general context, the present findings can be used for linking the attosecond and quantum information science (ATTOQUIS) towards the establishment of a roadmap for novel platforms of attosecond science and quantum technologies.  Contemporary quantum technologies face major difficulties in fault tolerant quantum computing with error correction, and focus instead on various shades of quantum simulation (Noisy Intermediate Scale Quantum devices \cite{preskill_quantum_2018}, analogue and digital Quantum Simulators \cite{georgescu_quantum_2014} and quantum annealers \cite{farhi_quantum_2001}). There is a clear need and quest for such systems that, without necessarily simulating dynamics of some quantum systems, can generate massive, controllable, robust, entangled and superpositions states. This will enable the use of these states for quantum communications \cite{gisin_quantum_2007} (e.g. to achieve transfer of information in a safer and quicker way), quantum metrology \cite{giovannetti_advances_2011}, sensing and diagnostics \cite{degen_quantum_2017} (e.g. to precisely measure phase shifts of light fields, or to diagnose quantum materials). To date, there are no existing platforms which bring processes at such short time-scales to quantum information science. ATTOQUIS can open the way for realizing a universal and firmly established tools to offer novel solutions and developments, i.e. a set of methods to generate massive entangled states and massive quantum superpositions for applications in quantum information science, having as final goal bringing them to quantum technologies.

\section{CONCLUSIONS}\label{Conclusion}
In this work, we investigated the quantum optics of strongly laser driven atoms. Using a fully quantized theoretical approach, we described the HHG and ATI processes and we showed how the conditioning on HHG and ATI processes can naturally lead to the generation of amplitude-shifted coherent-state superpositions. Additionally, we have investigated the parameters that can be used to control the quantum features of these states. This was experimentally confirmed by measuring the quantum features of the coherent-state superposition obtained after conditioning on HHG for different gas densities. We found that the coherent-state superposition changes from an optical ``cat'' to ``kitten'' state as the number of atoms participating in the harmonic generation process is reduced. We also show that this procedure can be used for the generation of high-photon-number coherent-state superpositions. This has been experimentally confirmed by recording a 9-photon shifted optical ``cat'' state. Finally, considering that the strong field laser-atom interaction is at the core of strong laser-field physics, it can be considered that our work builds the basis for the development of a new class of controllable high-photon-number non-classical light sources and for quantum optical studies of interactions induced in matter using laser intensities in the moderate and relativistic regions \cite{Nayak, Lamprou2021}.

\begin{acknowledgements}
We thank Jens Biegert, Ido Kaminer and Pascal Sali\`eres for enlightening discussions. We also thank I. Liontos, E. Skantzakis from FORTH and S. Karsch from Max Plank Institute for Quantum Optics for his assistance on maintaining the performance of the Ti:Sa laser system. ICFO group acknowledges support from ERC AdG
NOQIA, from Agencia Estatal de Investigación (the
R\&D project CEX2019-000910-S, funded by MCIN/
AEI/10.13039/501100011033, Plan National FIDEUA
PID2019-106901GB-I00, FPI, QUANTERA MAQS
PCI2019-111828-2, Proyectos de I+D+I “Retos Colaboración” RTC2019-007196-7) from Fundació Cellex, Fundació Mir-Puig, and from Generalitat de Catalunya
through the CERCA program, AGAUR Grant No. 2017
SGR 134, QuantumCAT U16-011424, co-funded by
ERDF Operational Program of Catalonia 2014-2020),
EU Horizon 2020 FET-OPEN OPTOLogic (Grant No
899794), and the National Science Centre, Poland
(Symfonia Grant No. 2016/20/W/ST4/00314), Marie
Skłodowska-Curie grant STREDCH No 101029393,
“La Caixa” Junior Leaders fellowships (ID100010434),
and EU Horizon 2020 under Marie Skłodowska-Curie
grant agreement No 847648 (LCF/BQ/PI19/11690013,
LCF/BQ/PI20/11760031, LCF/BQ/PR20/11770012).).
FORTH group acknowledges LASERLABEUROPE
(H2020-EU.1.4.1.2 Grant ID 654148), FORTH Synergy Grant AgiIDA (Grand No. 00133), the EU’s
H2020 framework programme for research and innovation under the NFFA-Europe-Pilot project (Grant
No. 101007417). J.R-D. acknowledges support from the Secretaria d'Universitats i Recerca del Departament d'Empresa i Coneixement de la Generalitat de Catalunya, as well as the European Social Fund (L'FSE inverteix en el teu futur)--FEDER. EP acknowledges support from Royal Society University Research
Fellowship URF\textbackslash R1\textbackslash 211390. P.S. acknowledges funding from the European Union’s Horizon 2020 research and innovation programme under the Marie Sklodowska-Curie grant agreement No 847517. A. S. M. acknowledges funding support from
the European Union’s Horizon 2020 research and innovation programme under the Marie Skłodowska-Curie grant agreement SSFI No.\ 887153. M. F. C. acknowledges support from the Guangdong Province Science and Technology Major Project Future functional materials under extreme conditions - 212019071820400001. P.T. group acknowledges LASERLABEUROPE V (H2020-EU.1.4.1.2 grant no.871124), FORTH Synergy Grant AgiIDA (grand no. 00133), the H2020 framework program for research and innovation under the NFFA-Europe-Pilot project (no. 101007417). ELI-ALPS is supported by the European Union and co-financed by the European Regional Development Fund (GINOP Grant No. 2.3.6-15-2015-00001). 
\end{acknowledgements}

\appendix
\renewcommand{\theequation}{\Alph{section}.\arabic{equation}}
\renewcommand{\thesection}{Appendix \Alph{section}}

\section{Quantum optical description of the laser-atom interaction: transformations and approximations}\label{SM1}
Our starting point is the time-dependent Schrödinger equation (TDSE) describing the interaction of the quantized electromagnetic field with a single electron
\begin{equation}\label{Org:Sch:eq}
	i\hbar \pdv{}{t}\ket{\tilde{\Psi}(t)} 
		= \hat{H}(t) \ket{\tilde{\Psi}(t)},
\end{equation}
where
\begin{equation}\label{H:def}
	\hat{H}(t) =
		\hat{H}_0 + \hat{H}_I + \hat{H}_f.
\end{equation}

Here, $\hat{H}_0 = \hat{\bf P}^2/2m + V(\hat{\bf R})$ is the Hamiltonian describing the electron bound to a potential $V(\hat{\bf R})$, $\hat{H}_I = -\text{e} \opE\cdot \opr$ is the dipole coupling that introduces the interaction between the electron and the field in the dipole approximation, and $\hat{H}_f$ is the electromagnetic free-field Hamiltonian. In the following, we will represent the electronic quadrature operators with capital letters ($\hat{X},\hat{P}$), while the photonic ones with lower-case letters ($\hat{x},\hat{p}$).

As we aim to describe laser/harmonic pulses of finite duration, we should consider in the free-field term $\hat{H}_f$ the full continuum spectrum of the electromagnetic field. Nevertheless, for the sake of simplicity, we write it as the sum of effective discrete modes containing the one obtained from the laser with frequency $\omega_L$ and its harmonics of frequencies $\omega_q=q\omega_L$, with $q=1,2,3,...$ up to the cut-off region of the spectrum. Concretely, we have 
\begin{equation}
	\hat{H}_f =
		\hbar \omega_L \hat{a}^\dagger \hat{a}
		+ \sum_{q=2}^\text{cutoff} \hbar q\omega_L \hat{b}_q^\dagger \hat{b}_q,
    \end{equation}
where $\hat{a}^\dagger$ ($\hat{a}$) and $\hat{b}_q^\dagger$ ($\hat{b}_q$) are the creation (annihilation) operators acting over the laser and the $q$th harmonic mode, respectively. Following the same idea, we model the laser electric field operator as
\begin{equation}\label{E:field:operator}
	\opE(t) =
		-i \hbar {\bf g}(\omega_L)f(t)
		\Big[
			 \big(\hat{a}^\dagger - \hat{a}\big)
			 + \sum_{q=2}^\text{cutoff}
			 	\sqrt{q} \big(\hat{b}_q^\dagger - \hat{b}_q\big)
		\Big].
\end{equation}

Here, we denote by ${\bf g}(\omega_L) \propto \sqrt{\omega_L/V_\text{eff}}$ the coefficient that enters into the expansion of the laser electric field modes and that depends on $V_\text{eff}$, which is the effective quantization volume \cite{QV1, QV2}. Thus, $\text{e}{\bf g}(\omega_L)$ encodes information about the polarization modes and has dimensions [m$^{-1}$s$^{-1}$]. Finally, $0 \leq f(t) \leq 1$ is a dimensionless function describing the pulse envelope.

At time $t=t_0$, we can describe the state of the system by $\ket{\Psi(t_0)} = \ket{\text{g}, \alpha_L, \Omega_H}$, that is, with the electron lying on the atomic ground state, the laser mode in a coherent-state and the harmonic modes in the vacuum state. Within this context, the first transformation we apply consists of moving to the interaction picture with respect to the electromagnetic field $\hat{H}_f$, i.e.,
\begin{equation}\label{Field:trans}
	\ket{\tilde{\Psi}(t)}
		= \exp[-i\hat{H}_ft] \ket{\Psi'(t)},
\end{equation}
so that Eq.~\eqref{Org:Sch:eq} reads
\begin{equation}
	i \hbar \pdv{}{t}\ket{\Psi'(t)}
		= \Big[ 
			   \hat{H}_0 - \text{e}\opE(t)\cdot \opr
		  \Big]
		  	   \ket{\Psi'(t)},
\end{equation}
where the laser electric field operator defined in Eq.~\eqref{E:field:operator} has an extra time dependence
\begin{equation}\label{E:op:time:}
	\begin{aligned}
		\opE(t)
			= &-i \hbar {\bf g}(\omega_L)f(t)
			 	\Big[
			 		 \big(\hat{a}^\dagger e^{i\omega_L t} - \hat{a}e^{-i\omega_L t}\big)
			 		\\&
			 		+ \sum_{q=2}^\text{cutoff}
			 		 \sqrt{q} \big(\hat{b}_q^\dagger e^{iq\omega_L t} - \hat{b}_qe^{-iq\omega_L t}\big)
				\Big].
	\end{aligned}
\end{equation}

The second transformation we apply consists of a displacement in the subspace of the driving laser field of a quantity $\alpha_L$, i.e.,
\begin{equation}\label{Disp:trans}
	\ket{\Psi'(t)}
		= \hat{D}(\alpha_L)\ket{\Psi(t)},
\end{equation}
where $\hat{D}(\alpha_L)$ is the optical displacement operator \cite{QV2}, acting over the laser mode. Recalling the following properties of this operator \cite{VogelWelsch}
\begin{equation}
	\hat{D}(\alpha)^\dagger \hat{D}(\alpha)=\mathbbm{1},
\end{equation}
\begin{equation}
	\hat{D}(\alpha) \hat{a} \hat{D}^\dagger(\alpha)
		= \hat{a} - \alpha,
\end{equation}
its introduction in our equations has two mutually related consequences: it sets the initial state of the laser mode to a vacuum state $\Omega_L$, and transforms our TDSE into
\begin{equation}\label{Sc:Sch:eq}
	\begin{aligned}
	i \hbar \pdv{}{t}\ket{\Psi(t)}
		&= \Big[ 
			   \hat{H}_0 - \text{e}{\bf E}_L(t)\cdot \opr -\text{e}\opE_Q(t)\cdot \opr
		  \Big]
		  	   \ket{\Psi(t)}\\
		&= \Big[
				\hat{H}_\text{sc} -\text{e}\opE_Q(t)\cdot \opr
			\Big] \ket{\Psi(t)} .
	\end{aligned}
\end{equation}

Here, ${\bf E}_L(t)$ accounts for the classical electric field part of the laser pulse
\begin{equation}
	{\bf E}_L(t)
		= -i \hbar {\bf g}(\omega_L)  f(t)
		  \Big[
		  	   \alpha_L^* e^{i\omega_L t}
		  	   - \alpha_L e^{-i\omega_L t}
		  \Big],
\end{equation}
so $\hat{H}_\text{sc}$ represents the semiclassical part of our Hamiltonian \cite{Lewenstein1994}. On the other hand, $\opE_Q(t)$ is the quantum correction term defined as in Eq.~\eqref{E:op:time:}.

Lastly, we move to the interaction picture with respect to the semiclassical Hamiltonian $\hat{H}_\text{sc}$
\begin{equation}
	\ket{\Psi(t)}
		= \mathcal{T} 
			\exp[-i\int^t_{t_0} \dd t' \hat{H}_\text{sc}(t')/\hbar]
			\ket{\psi(t)},
\end{equation}
where $\mathcal{T}$ is the time-ordering operator. This last transformation leads us to the final form of our TDSE, which we will use throughout this manuscript, i.e.
\begin{equation}\label{Final:Sch:Eq}
	i\hbar \pdv{}{t} \ket{\psi(t)}
		= -\text{e} \opE_Q(t) \cdot \opr_H(t) \ket{\psi(t)},
\end{equation} 
where $\text{e}\opr_H(t)$ denotes the time-dependent dipole operator in the considered semi-classical interaction picture, acting exclusively on the electronic degrees of freedom. This evolution drives the dynamics of the field and the electron, which may end up in the ground or continuum states. On the other hand, we consider, the electron will rarely end up in a bound-excited state.

\section{Quantum optical description of high-harmonic generation}\label{SM2}
In the HHG process, the electron gets first transferred to the continuum via tunneling ionization due to the strong laser field we are applying and, later on, it recombines with the parent ion that was left behind, ending up again in the ground state of the system. Therefore, in order to get information about the HHG photonic quantum state, we condition Eq.~\eqref{Final:Sch:Eq} onto the atomic ground state $\ket{\text{g}}$, i.e.,
\begin{equation}\label{HHG:int:Sch:eq}
	i \hbar \pdv{}{t} \braket{\text{g}}{\psi(t)}
		= -\opE_Q(t) \cdot \mel{\text{g}}{\text{e}\opr_H(t)}{\psi(t)}.
\end{equation}

Defining the identity operator as
\begin{equation}\label{Id:op}
	\mathbbm{1}
		= \dyad{\text{g}}
		  + \sum_{\phi_b}\dyad{\phi_b}
		  + \int \dd \phi_c \dyad{\phi_c},
\end{equation}
where we denote with the discrete sum the set of atomic bound excited states, and with the integral the set of continuum states, we introduce it in Eq.~\eqref{HHG:int:Sch:eq} to get
\begin{equation}\label{HHG:gen:Sch:eq}
	\begin{aligned}
	i \hbar \pdv{}{t} \braket{\text{g}}{\psi(t)}
		= & -\opE_Q(t) \cdot
		 \Big[
			\meld_H(t)\braket{\text{g}}{\psi(t)}\\
			 &\hspace{1.5cm}
			 + \sum_{\phi_b} \meld_H(\phi_b,t) \braket{\phi_b}{\psi(t)}\\
			 &\hspace{1.5cm}
			    + \int \dd\phi_c \meld_H(\phi_c,t)\braket{\phi_c}{\psi(t)}
		\Big].
	\end{aligned}
\end{equation}

In this last expression, we denote with $\meld_H(t) = \mel{\text{g}}{\text{e}\opr_H(t)}{\text{g}}$ the quantum averaged time-dependent dipole moment and with $\meld_H(\phi_k, t)= \mel{\text{g}}{\text{e}\opr_H(t)}{\phi_k}$ the matrix element between the ground state and state $\ket{\phi_k}$, where $k$ can take values $b$ or $c$ depending on whether the state belongs to the bound excited states or to the continuum region of the spectrum, respectively. Each of these terms is multiplied by the probability amplitude of finding the electron either in the ground state, in another excited bound state or in an excited continuum state. In the first attempt to solve the problem, we will assume that these two last terms are very small in comparison to the first one, which is a fair assumption as the electron hardly remains in an excited bound/continuum state \cite {Symphony, Lewenstein1994} at the end of the pulse. Therefore, our TDSE adopts the following form
\begin{equation}\label{HHG:Final:Sch:eq}
	i \hbar \pdv{}{t} \ket{\Phi(t)}
		=  -\opE_Q(t) \cdot \meld_H(t)\ket{\Phi(t)},
\end{equation}
where $\ket{\Phi(t)} = \braket{\text{g}}{\psi(t)}$. Here, $\meld_H(t)$ can be easily calculated by numerically solving the TDSE, or by means of the strong-field approximation (SFA) theory \cite{Symphony, Lewenstein1994, ScrinziBook}. Whatever the method used, this equation can be easily solved as it is written as a linear combination of photon creation and annihilation operators for the different modes considered in the problem. This has a natural implication, and is that the final solution is given by a product state of all the modes participating in the process,
\begin{equation}
	\ket{\Phi(t)}
		= \ket{\Phi_{q=1}(t)} 
		  \otimes \ket{\Phi_{q=2}(t)}
		  \otimes \dots
		  \otimes \ket{\Phi_{q= \text{cutoff}}(t)},
\end{equation}
so we can solve the equation for a given $q$ and then generalize the result to the rest. Thus, the single mode version of Eq.~\eqref{HHG:Final:Sch:eq} which we will now deal with is
\begin{equation}\label{HHG:1mode:eq}
	\begin{aligned}
	i\hbar \pdv{}{t} \ket{\Phi_q(t)}
		&=  -\opE_q(t) \cdot \meld_H(t)\ket{\Phi(t)}
		= \hat{H}_q(t) \ket{\Phi(t)},
	\end{aligned}
\end{equation}
where
\begin{equation}	\label{E:field:1mode}
	\opE_q(t) = -i \hbar {\bf g}(\omega_L) f(t) \sqrt{q}
				\Big[
					 \hat{b}^\dagger_q e^{iq\omega_L t}
					 - \hat{b}_q e^{-iq\omega_L t}
				\Big].
\end{equation}

In general, we can write the solution to this equation as \cite{Tannor}
\begin{equation}
	\ket{\Phi_q(t)}
		= \hat{U}_q(t,t_0) \ket{\Phi_q(t_0)},
\end{equation}
where $\hat{U}(t,t_0)$ is our time-evolution operator. Furthermore, we can split our time interval in $N$ steps of size $\Delta t$, which is typically defined to be inversely proportional to $N$, such that we can write this operator as
\begin{equation}\label{Time:Ev:Op}
	\hat{U}_q(t,t_0) = 
	    \lim_{N\to\infty} 
		\prod^{N-1}_{i=0}
		\hat{U}_q(t_{i+1}, t_i),
\end{equation}
where we identify $t_N = t$. Therefore, we can write each of the unitary operators appearing in the previous product as
\begin{equation}
	\hat{U}_q(t_{i+1}, t_i)
		= \exp[-i \hat{H}_q(t_{i+1}) \Delta t/ \hbar].
\end{equation}

Let us take a closer look to the commutation relation between $\hat{H}_q(t)$ defined at two different times $t$ and $t'$
\begin{equation}\label{Comm:rel}
    \begin{aligned}
	i\comm{\hat{H}_q(t)}{\hat{H}_q(t')}
		= &- 2q\hbar^2 f(t)f(t') 
		\\& \times
		\Big({\bf g}(\omega_L) \cdot \meld_H(t)\Big)
			 \Big({\bf g}(\omega_L) \cdot \meld_H(t')\Big)
		\\& \times
		\sin(q\omega_L(t-t')) \mathbbm{1}.
	\end{aligned}
\end{equation}

As we can see, this term is a function proportional to the identity operator, something that favours the implementation of the Baker-Campbell-Hausdorff (BCH) formula \cite{QV3}, i.e.,
\begin{equation}
	e^{\hat{X}} e^{\hat{Y}} = e^{\hat{Z}}
\end{equation} 
where
\begin{equation}\label{Z:Def}
	\hat{Z} =
		\hat{X} + \hat{Y}
		+ \dfrac12 \comm{\hat{X}}{\hat{Y}}
		+ \dfrac{1}{12} \comm{\hat{X}}{\comm{\hat{X}}{\hat{Y}}}
		+ \dots,
\end{equation}
to join all the exponential operators in Eq.~\eqref{Time:Ev:Op}, as we only have to keep the first three terms in the right hand side of Eq.~\eqref{Z:Def} since all the other terms commute. Notice that each time we join two consecutive operators, we get an extra exponential term from the commutation relation in Eq.~\eqref{Comm:rel}. The exponent of such term adopts the following form
\begin{equation}\label{BCH:pref}
	i\varphi_q(t)
		= -\dfrac{i}{2}
		    \sum^{N-1}_{j=1}\sum^j_{i=0}
			\comm{\hat{H}_q(t_j)}{\hat{H}_q(t_i)}\Delta t^2/\hbar^2
\end{equation}
and the final time-evolution operator reads 
\begin{equation}\label{U:q}
	\hat{U}_q(t,t_0)
		= \lim_{N\to\infty}
		  \exp[-i 
			   \bigg(
			   		\sum_{i=0}^{N-1} \hat{H}_q(t_i)
			   	\bigg)\Delta t/\hbar
			   ]
		   e^{i\varphi_q(t)},
\end{equation}
where in the exponential operator term we naturally recover the definition of an integral. Thus, according to the definition of $\hat{H}_q(t)$ given in Eq.~\eqref{HHG:1mode:eq}, the previous unitary operator can be written, for the case of the fundamental mode ($q=1$), as
\begin{equation}\label{Evolution:L}
    \hat{U}_L(t,t_0)
        = \exp[
                \delta \alpha_L \hat{a}^\dagger
                -\delta \alpha_L^* \hat{a}
            ]
            e^{i\varphi_L(t)}
\end{equation}
which is a displacement in the photonic phase space of a quantity $\delta \alpha_L$ defined by
\begin{equation}\label{def:dalpha}
    \delta \alpha_L(t)
        = {\bf g}(\omega_L) \cdot 
          \int_{t_0}^t\dd \tau\ 
            f(\tau) {\bf d}_H(\tau) e^{i\omega_L \tau}. 
\end{equation}

Therefore, incorporating the action of Eq.~\eqref{U:q} over the harmonic modes, we finally get the final quantum optical HHG state
\begin{equation}\label{HHG:qstate}
	\begin{aligned}
	\ket{\Phi(t)}
		= \ &e^{i\varphi_L(t)}\ket{(\alpha_L + \delta\alpha_L)e^{-i\omega_L t}}
			\otimes e^{i\varphi_2(t)}\ket{\beta_2 e^{-i2\omega_L t}}
			\\ &
			\otimes \dots
			\otimes e^{i\varphi_q(t)}\ket{\beta_q e^{-iq\omega_L t}}
			\otimes \dots,
	\end{aligned}
\end{equation}
where we have returned to the original photonic frame of reference, that is, we have undone the initial transformations depicted in Eqs.~\eqref{Field:trans} and \eqref{Disp:trans}. Note that here the $i\varphi_q(t)$ are defined as in Eq.~\eqref{BCH:pref} once the limit $N \to \infty$ has been considered. Similarly to the $\delta \alpha_L$, the $\beta_q$ terms are defined as
\begin{equation}
    \beta_q(t)
        = \sqrt{q}\ {\bf g}(\omega_L) \cdot 
          \int^t_{t_0}\dd \tau\ 
            f(\tau) {\bf d}_H(\tau) e^{iq\omega_L \tau}. 
\end{equation}

The results obtained until now are valid for the single-atom case. For the $N$-atomic case, assuming that each atom contributes to the HHG process coherently in a phase matched way, the definitions of $\delta \alpha_L$ and $\beta_q$ are reformulated as,
\begin{equation}\label{def:alpha:multi}
	\delta \alpha_L(t)
		= N {\bf g}(\omega_L) \cdot 
          \int^t_{t_0}\dd \tau\ 
            f(\tau) {\bf d}_H(\tau) e^{i\omega_L \tau}
\end{equation} 
\begin{equation}\label{def:beta:multi}
	\beta_q(t)
		= N \sqrt{q}\ {\bf g}(\omega_L) \cdot 
          \int^t_{t_0}\dd \tau\ 
            f(\tau) {\bf d}_H(\tau) e^{iq\omega_L \tau}.
\end{equation}

Note that in this case the $N$-atomic wavefunction will be affected by an overall phase coming from the BCH relation, that does not affect the phase matching conditions which are solely determined by the phase of the generated coherent-states. In order to give a physical meaning to $\delta \alpha_L$ and $\beta_q$ within the electron recollision picture, we will use the strong-field approximation theory to provide a solution to the integrals in Eqs.~\eqref{def:alpha:multi} and \eqref{def:beta:multi}. According to the SFA, it can be shown \cite{Lewenstein1994} that the mean value of the dipole operator ${\bf d}_H(t)$ reads
\begin{equation}\label{SFA:dipole}
    \begin{aligned}
    {\bf d}_H(t) =
        i \int^t_{t_0} \! \! \dd t' \!\! \int &\dd {\bf v} \
            {\bf d}^*\Big({\bf p} -\frac{e}{c}\boldsymbol{A}_L(t') \Big)
             e^{-i S({\bf p},t,t')}
             \\ & \times
             \boldsymbol{E}_L(t')
             {\bf d} \Big({\bf p} -\frac{e}{c}\boldsymbol{A}_L(t')
             \Big)
            + \ \text{c.c.},
    \end{aligned}
\end{equation}
where $\boldsymbol{A}_L(t)$ is the vector potential of the laser field defined as $\boldsymbol{E}_L(t) = -(1/c) \pdv*{\boldsymbol{A}_L(t)}{t}$, ${\bf p} = {\bf v} + (e/c)\boldsymbol{A}_L(t)$ the canonical momentum whereas ${\bf v}$ the electron's kinetic momentum, ${\bf d} ({\bf p} -(e/c)\boldsymbol{A}_L(t'))$ is the matrix element of the dipole operator between the atomic ground state and the continuum state $\ket{{\bf p} -(e/c)\boldsymbol{A}_L(t')}$, and $S({\bf p},t,t')$ is the semiclassical action given by
\begin{equation}\label{Semiclass:action}
    \begin{aligned}
    S({\bf p},t,t')
        = \dfrac12 \int^t_{t'}\dd \tau &
            \Big[
                {\bf p} -\frac{e}{c}\boldsymbol{A}_L(\tau)
            \Big]^2
            + I_p(t-t'),
    \end{aligned}
\end{equation}
where $I_p$ is the ionization potential. 

For the sake of simplicity, we will assume that the used laser field consists of a monochromatic field of frequency $\omega_L$, so that we can set $f(t) = 1$ in Eqs.~\eqref{def:alpha:multi} and \eqref{def:beta:multi} which now read
\begin{equation}\label{def:alpha:mono}
	\delta \alpha_L(t)
		= N {\bf g}(\omega_L) \cdot 
          \int^t_{t_0}\dd \tau\ 
            {\bf d}_H(\tau) e^{i\omega_L \tau}
\end{equation} 
\begin{equation}\label{def:beta:mono}
	\beta_q(t)
		= N \sqrt{q}\ {\bf g}(\omega_L) \cdot 
          \int^t_{t_0}\dd \tau\ 
            {\bf d}_H(\tau) e^{iq\omega_L \tau}.
\end{equation}

The semiclassical action shown in Eq.~\eqref{Semiclass:action} is a highly oscillating function which leads to a high oscillating exponent in Eq.~\eqref{SFA:dipole}, and allows for a solution to the triple integration appearing in Eqs.~\eqref{def:alpha:mono}, \eqref{def:beta:mono} by means of the saddle-point approximation. Therefore, the integrals in Eqs.~\eqref{def:alpha:mono},~\eqref{def:beta:mono} are completely characterized by the saddle-points determined by the set of variables $({\bf p_s}, t_r, t_i)$ fixed by the following three equations that have been extensively studied in the past within the context of the semiclassical three-step model \cite{Symphony, Lewenstein1994},     
\begin{align}
    &\dfrac{[{\bf p_s} -\frac{e}{c}\boldsymbol{A}_L(t_i)]^2}{2} + I_p = 0, \label{saddle:point:1}\\
    &\int^{t_r}_{t_i}\dd \tau \Big[{\bf p_s} -\frac{e}{c}\boldsymbol{A}_L(\tau)\Big] = 0,\label{saddle:point:2}\\
    &\dfrac{[{\bf p_s} -\frac{e}{c}\boldsymbol{A}_L(t_r)]^2}{2} + I_p = q\omega_L.\label{saddle:point:3}
\end{align}

In brief terms, the above equations define the three-steps of the recollision process: \eqref{saddle:point:1} defines the ionization time $t_i$, \eqref{saddle:point:2} the electron's return to the parent ion, and \eqref{saddle:point:3} the recombination time $t_r$ associated with the generation of high harmonics with frequencies $q\omega_L > I_p$. On the one hand, these equations imply that the shift $\delta \alpha_L$ of the coherent-state is directly related to the electron ionization and acceleration processes. On the other hand, they also show that the well-known features of the HHG process are transferred to the coherent-states of the harmonic field, that is, the $\beta_q$'s contain information about the spectral phase and amplitude distribution of the emitted harmonics. In fact, this can be shown by calculating the spectrum of the generated harmonics, which can be obtained from their energy $\langle \hat{H}_f\rangle_\text{em} = \sum_q \hbar \omega_q n_q$. In this expression, $n_q$ is the number of photons at frequency $\omega_q = q\omega_L$ which, according to Eq.~\eqref{def:beta:multi}, is given by
\begin{equation}\label{Nq}
    n_q =
        N^2 \lvert{\bf g}(\omega_q)\cdot{\bf d}_H(q\omega_L)\rvert^2.
\end{equation}

To obtain Eq.~\eqref{Nq}, we have sent the integration limits to $\pm \infty$, implying that the electric field is introduced at $t_0 = -\infty$ and lasts until $t = + \infty$, so that the integral appearing in Eq.~\eqref{def:beta:mono} represents the Fourier transform of the mean-valued dipole ${\bf d}_H(q\omega_L)$. Considering all possible frequencies, its summation can be rewritten as an integral, and the energy of the emitted harmonics reads
\begin{equation}\label{IntegrtedHarmEnergy}
    E_\text{em}
        =  \dfrac{V_\text{eff}}{(2\pi c)^3}
        \int \dd \Omega \ \dd \omega \ 
        N^2\omega^3 \lvert{\bf g}(\omega)\cdot
        {\bf d}_H(\omega)\rvert^2,
\end{equation}
where $\dd \Omega$ represents the infinitesimal solid angle element. Substituting the definition of ${\bf g}(\omega_L$) into Eq. \eqref{IntegrtedHarmEnergy}, we find for its integrand
\begin{equation}\label{HHG:spectrum}
    \mathcal{E}_\text{HHG}(\omega_q)
        \propto 
        N^2 \omega_q^4 |{\bf d}_H(\omega_q)|^2,
\end{equation} 
which corresponds to the expression of the HHG spectrum obtained by the semiclassical theory \cite{Symphony, Lewenstein1994}.

\section{Analysis of the coherent shift in the fundamental mode}\label{SM3}
As mentioned in the text and explicitly developed in \ref{SM2}, the shift in the fundamental mode $\delta \alpha_L$ is related to the absorbed part of the driving field that is necessary for generating the harmonic photons, its properties can be related to the exchange of photons during the interaction. In particular, we are interested in the probability of absorbing $n$ photons during the ionization and acceleration processes. For that reason, we consider a coherent-state $\ket{\delta\alpha_L(t,t_0)}$ and compute the probability distribution of having $n$ photons on it
\begin{equation}
    \begin{aligned}
    P_n(t,t_0)
        &=
        \lvert\braket{n}{\delta \alpha_L(t,t_0)}\rvert^2\\
        &= \dfrac{\lvert\delta \alpha_L(t,t_0)\rvert^{2n}}{n!}
          e^{-\lvert\delta \alpha_L(t,t_0)\rvert^2}.
    \end{aligned}
\end{equation}

We note that this quantity is related to the probability of absorbing $n$ photons during the ionization and acceleration processes. We further introduce the average probability of having $n$ photons in the above coherent-state within a cycle of the field that starts at $t_0$ and finishes at time $T$ as,
\begin{equation}\label{Time:average}
    \tilde{P}_n
        =\dfrac{1}{T - t_0} 
        \int^T_{t_0} \dd t 
            P_n(t,t_0).
\end{equation}

\begin{figure}[ht!]
	\includegraphics[width=1\columnwidth]{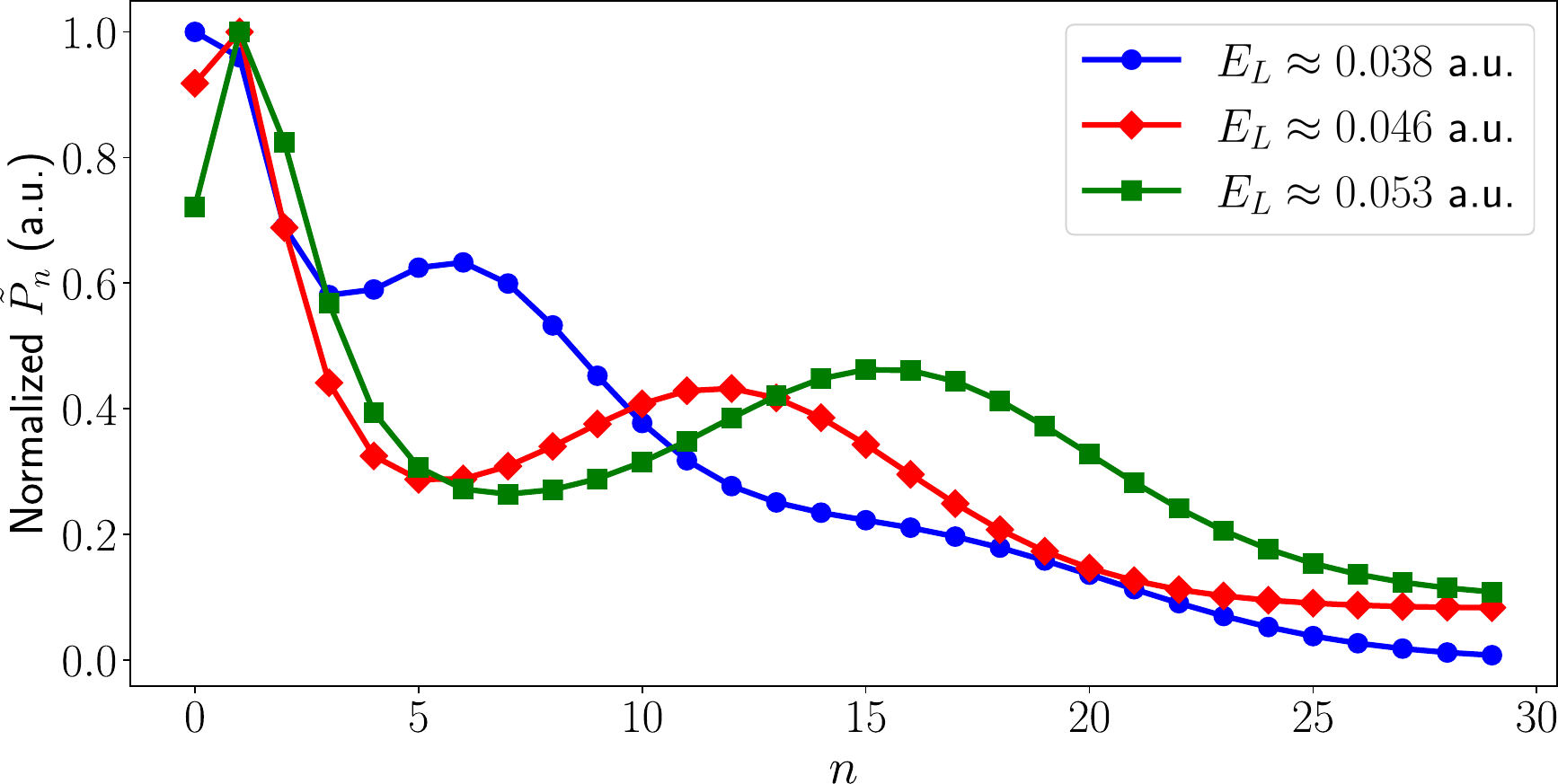}
	\caption{Probability of having $n$ photons in the coherent-state $\lvert\delta\alpha_L\rangle$ averaged in time. Normalized average probability of a single atom to absorb $n$ photons for three different electric field amplitudes, $E_L\approx 0.053$ a.u. (green squared-dotted curve), $E_L \approx 0.046$ a.u. (red rhomboid-dotted curve) and $E_L\approx 0.038$ a.u. (blue round-dotted curve)) of the driving field (in atomic units). The results have been obtained by integrating over one cycle of a gaussian shaped pulse with central wavelength $\lambda_L = 800$ nm.} 
	\label{SM3Fig1}
\end{figure}

The numerical results obtained from this calculation are shown in Fig.~\ref{SM3Fig1} for three different intensities of the driving field. As we can see, for each of the curves we get a local maximum in the probability which shifts to bigger values of the number of photons $n$ as the intensity of the field increases. This is consistent with the harmonic plateau structure obtained for the HHG spectrum. As the intensity increases the harmonic cutoff is extended to higher photon number values and, in consequence, photons of higher frequency are achievable through the HHG process. Thus, given that for generating a photon of frequency $n\omega_L$ a number $n$ of IR photons need to be absorbed, then in order to get a plateau structure for the harmonic spectrum the probability of absorbing IR photons should increase as we move towards the harmonic cutoff, reaching a maximum at this point and decreasing afterwards. To check this, we look at the value of $n$ for which we find a local maximum in the probability (the maximum obtained for $n>2$) for each of the considered intensities. In particular, in Fig.~\ref{SM3Fig1} these maxima are placed at $n_\text{cutoff}\approx 7, 12$ and $15$ from the lowest to the highest intensity respectively, which are in agreement with the cutoffs given by the maximum kinetic energy that an electron can get in the HHG process with the corresponding intensities (the theoretical values for the cutoff are $n_\text{th} = 7.71, 11.3$ and $15.0$ respectively). Note that in comparison to Eq.~\eqref{HHG:spectrum}, here we do not obtain a multi-peak structure involving only the odd harmonics. This is because $\delta \alpha$ describes the amount of IR photons absorbed during the ionization and acceleration processes affecting the fundamental laser mode, which later on will be distributed along the generated harmonics.

\section{Conditioning onto HHG: Generation of Schrödinger optical ``kitten'' and ``cat'' states}\label{SM4}
As mentioned in the main text, the time-evolved state obtained after conditioning the electron state to be end up in the ground state of the system is given by
\begin{equation}\label{eq:HHG:qstate:}
	\begin{aligned}
	\ket{\Phi(t)}
		= \ &e^{i\varphi_L(t)}\ket{(\alpha_L + \delta\alpha_L)e^{-i\omega_L t}}
			\otimes e^{i\varphi_2(t)}\ket{\beta_2 e^{-i2\omega_L t}}
			\\ &
			\otimes \dots
			\otimes e^{i\varphi_q(t)}\ket{\beta_q e^{-iq\omega_L t}}
			\otimes \dots,
	\end{aligned}
\end{equation}

The key action for the creating non-classical states of light is the post-selection of the coherent shifted IR state over the part that includes, at least, one harmonic photon. It was shown in \cite{Stammer2021} that, conditioning the harmonic modes to be found in the state $\bigotimes^\text{cutoff}_{q=2} \ket{\beta_q}$ and considering very high values of the harmonic cutoff, the final quantum optical state of the infrared mode is given by (up to normalization) by
\begin{equation}\label{eq:HHG:catstate:}
    \ket{\Phi_\text{post}}
        \approx \ket{\alpha_L + \delta\alpha_L} 
            - \braket{\alpha_L}{\alpha_L + \delta\alpha_L}
                \ket{\alpha_L}.
\end{equation}

In the following we will explicitly develop the different situations studied in the main text that lead us to the generation of Schrödinger optical ``kitten'' and cat states.

\subsection{Obtaining a ``kitten'' state}
The kitten state is obtained in the limit when $\braket{\alpha_L}{\alpha_L + \delta\alpha_L} \to 1$, which corresponds to the limit where $\lvert \delta \alpha_L\rvert \to 0$. This is valid whenever $\delta \alpha$ adds a depletion to the initial coherent-state, a condition that is verified when the phase of $\delta \alpha$ and $\alpha$, which we denote here as $\theta_\delta$ and $\theta_\alpha$ respectively, satisfy 
\begin{equation}
    \dfrac{\pi}{2} 
        + \arcsin(\dfrac{\lvert\delta \alpha_L\rvert}{2 \lvert \alpha_L \rvert}) + \theta_\alpha 
            < \theta_\delta 
                < \dfrac{3\pi}{2} 
                    - \arcsin(\dfrac{\lvert\delta \alpha_L\rvert}{2 \lvert \alpha_L \rvert}) + \theta_\alpha.
\end{equation}

In order to work in regimes of vanishing $\lvert\delta \alpha_L\rvert$, we will consider an expansion of the postprocessed state presented in Eq.~\eqref{eq:HHG:catstate:} in terms of powers of $\lvert\delta \alpha_L\rvert$. With that purpose, we first write our shifted coherent-state $\ket{\alpha_L + \delta \alpha_L}$ as
\begin{equation}\label{App:Derivative:depleted}
    \begin{aligned}
    \ket{\alpha_L + \delta \alpha_L}
        &= \hat{D}(\alpha_L)e^{\frac12(\alpha_L^*\delta \alpha_L - \alpha \delta\alpha_L^*)}
            \hat{D}(\delta \alpha_L)\ket{0},
    \end{aligned}
\end{equation}
where we have considered the following property of the displacement operator
\begin{equation}
    \hat{D}(\alpha_L+\delta\alpha_L) =
    e^{\frac12(\alpha^*\delta \alpha_L - \alpha_L \delta\alpha_L^*)}
    \hat{D}(\alpha_L)\hat{D}(\delta \alpha_L).
\end{equation}

Introducing here the definition of the displacement operator $D(\alpha)$ and in particular its polynomial expansion
\begin{equation}
    \begin{aligned}
    \hat{D}(\delta \alpha_L) 
        &= \exp[\delta\alpha_L \hat{a}^\dagger - \delta\alpha_L^* \hat{a}]
        \\&
        = \sum_{n=0}^{\infty} \dfrac{(\delta \alpha_L \hat{a}^\dagger - \delta \alpha_L\hat{a})^n}{n!}\\
        &= \sum_{n=0}^{\infty} \lvert\delta \alpha_L\rvert^n
        \dfrac{(e^{i\theta_\delta} \hat{a}^\dagger - e^{-i\theta_\delta}\hat{a})^n}{n!},
    \end{aligned}
\end{equation}
which, introduced in Eq.~\eqref{App:Derivative:depleted}, leads to the desired polynomial expansion in $|\delta \alpha|$
\begin{equation}\label{App:CalcDeriv:depleted}
    \begin{aligned}
    \ket{\alpha_L + \delta \alpha_L}
        &= \hat{D}(\alpha_L)e^{\frac12(\alpha_L^*\delta \alpha_L - \alpha_L \delta\alpha_L^*)}
        \\ & \quad\times
        \sum_{n=0}^{\infty} 
        \lvert\delta \alpha_L\rvert^n
            \dfrac{(e^{i\theta_\delta} \hat{a}^\dagger - e^{-i\theta_\delta}\hat{a})^n}{n!}
                \ket{0},
    \end{aligned}
\end{equation}
and whose scalar product with $\ket{\alpha_L}$ is given by
\begin{equation}\label{App:braket}
    \begin{aligned}
    \braket{\alpha}{\alpha_L+\delta \alpha_L}
        &= e^{\frac12(\alpha_L^*\delta \alpha_L - \alpha_L \delta\alpha_L^*)}
            \braket{0}{\delta\alpha_L}\\
        &= e^{\frac12(\alpha_L^*\delta \alpha_L - \alpha_L \delta\alpha_L^*)}
            e^{-\lvert\delta\alpha_L\rvert^2/2}\\
        &=e^{\frac12(\alpha_L^*\delta \alpha_L - \alpha_L \delta\alpha_L^*)}
            \sum_{n=0}^\infty 
            \dfrac{1}{2^{n}}
            \dfrac{\lvert \delta\alpha_L\rvert^{2n}}{n!}.
    \end{aligned}
\end{equation}

Combining Eqs.~\eqref{App:CalcDeriv:depleted} and \eqref{App:braket} with Eq.~\eqref{eq:HHG:catstate:} we then get
\begin{equation}
    \begin{aligned}
     \ket{\Phi_\text{post}} 
        &= e^{\frac12(\alpha^*\delta \alpha_L - \alpha_L \delta\alpha_L^*)}
            \hat{D}(\alpha_L)
            \\& \hspace{-0.5cm}\times
            \sum_{n=1}^\infty 
            \bigg(
                \lvert\delta \alpha_L\rvert^n
                \dfrac{(e^{i\theta_\delta} \hat{a}^\dagger - e^{-i\theta_\delta}\hat{a})^n}{n!}
                -
                \dfrac{1}{2^{n}}
                \dfrac{\lvert \delta\alpha_L\rvert^{2n}}{n!}
            \bigg)
            \ket{0},
    \end{aligned}
\end{equation}
where we start the sum at $n=1$ because the $n=0$ term cancels due to the equal contribution of the two terms in the difference. Thus, the previous difference leads to
\begin{equation}
    \begin{aligned}
    \ket{\Phi_\text{post}} 
        &=e^{\frac12(\alpha_L^*\delta \alpha_L - \alpha_L \delta\alpha_L^*)}
            \hat{D}(\alpha_L)
            \\&\quad\times
            \Big(
                \delta\alpha_L \hat{a}^\dagger \ket{0} + \mathcal{O}\big(|\delta\alpha_L|^2\big)\Big),
    \end{aligned}
\end{equation}
which up to first order in $|\delta \alpha|$ corresponds with the definition of a displaced Fock state. Furthermore, we note that the photon number probability distribution of this state is given by
\begin{equation}
		P(n) 
			= \Big\lvert \dfrac{n}{\alpha_L}-\alpha_L^*\Big\rvert^2
			  \dfrac{\lvert\alpha_L\rvert^{2n}}{n!} e^{-\lvert\alpha_L\rvert^2}
\end{equation}
and whose Wigner function \cite{QV3} is characterized by
\begin{equation}
	\begin{aligned}
	W(\beta)
		&= \dfrac{2}{\pi}\tr\big(
				   \hat{D}(\beta)  \hat{\Pi} \hat{D}(-\beta)\dyad{\Phi_\text{post}}
			   \big)
		\\&
		= \dfrac{2}{\pi}(4\lvert\beta - \alpha_L\rvert^2 -1)
		   e^{\lvert\beta - \alpha_L\rvert^2/2}.
	\end{aligned}
\end{equation}

For obtaining this expression, we have used the Wigner function definition of ref.~\cite{Royer1977}, where $\hat{\Pi}$ denotes the parity operator, whose action over the displacement operator is given by $D(-\alpha) = \Pi D(\alpha) \Pi$.

\subsection{Obtaining a genuine ``cat'' state}
On the other hand, in the regime where $0 < \braket{\alpha_L + \delta \alpha_L}{\alpha_L} < 1$, we obtain a genuine ``cat'' state (shown in Eq.~\eqref{eq:HHG:qstate:}) with photon number probability distribution
\begin{equation}
	\begin{aligned}
	P(n)
		=& \dfrac{1}{N_\text{cat}}\Big\lvert(\alpha_L + \delta \alpha_L)^n e^{-\lvert\alpha_L + \delta \alpha_L\rvert^2/2}
		\\&
		 - \braket{\alpha_L}{\alpha_L + \delta \alpha_L}
		 \alpha^n e^{-\lvert\alpha_L\rvert^2/2}\Big\rvert^2,
	\end{aligned}
\end{equation}
and Wigner function
\begin{equation}
	\begin{aligned}
	W(\beta)
		=\ &\dfrac{2}{\pi N_\text{cat}}
		 \Big[ e^{-2\lvert\beta - \alpha_L - \delta\alpha_L\rvert^2}
		 + e^{-\lvert\delta \alpha_L\rvert^2}e^{-2\lvert\beta - \alpha_L\rvert^2}
		 \\
		 & \hspace{1cm}
		 - \big(
		 		e^{2(\beta - \alpha_L)\delta \alpha_L^*}
		 		+ e^{2(\beta - \alpha_L)^*\delta \alpha_L}
		 	\big)
		 	\\
		 	&\hspace{1cm}\quad\times
		 	e^{-\lvert\delta \alpha_L\rvert^2}e^{-2\lvert\beta - \alpha_L\rvert^2}
		 	\Big],
	\end{aligned}
\end{equation}
where $N_\text{cat} = 1 - e^{-\lvert\delta \alpha_L\rvert^2}$ is the normalization factor for Eq.~\eqref{eq:HHG:catstate:}.

We finally note that in the regime where $\lvert \delta \alpha_L\rvert$ becomes large enough so that $\braket{\alpha_L}{\alpha_L + \delta \alpha_L}\to 0$, we get a coherent shifted state with photon number probability distribution given by a poissonian
\begin{equation}
    P(n) 
        = e^{-\lvert \alpha_L + \delta \alpha_L\rvert^2}
            \dfrac{\lvert\alpha_L + \delta \alpha_L\lvert^{2n}}{n!},
\end{equation}
and Wigner function
\begin{equation}
    	W(\beta)
		=\dfrac{2}{\pi N_\text{cat}}
		  e^{-2\lvert\beta - \alpha_L - \delta\alpha_L\rvert^2}.
\end{equation}

\section{Quantum optical description of above-threshold ionization}\label{SM5}
We showed in the main text that, under the strong-field approximations and within the single active electron scenario, the conditioned to ATI quantum optical state is given by \begin{equation}\label{eq:ATI:SingleState:}
    \ket{\Tilde{\Phi}(\textbf{v},t)}
        \approx i\hbar \sum_{j=0}^{\mathcal{N}-1}
        \int^{t_{j+1}}_{t_j} \! \! \! \! \! \dd t'\
        \opE_L(t') \cdot \meld^*_H({\bf v},t')
        \ket{(j+1)\Delta},
\end{equation}
where $\ket{\Phi(t)}$ is given by Eq.~\eqref{eq:HHG:qstate:} before going back to the laboratory frame (which we get by setting $\alpha_L$ = 0 in the mentioned state).

In the main text we consider two possible strategies for deriving the reduced density matrix of the electromagnetic field after conditioning to ATI:
\begin{enumerate}[{\bf I)}]
    \item We can condition on ATI electrons with a specific outgoing direction and kinetic momentum, ${\bf v}$. In this approach the reduced density matrix of the system is given by,
    \begin{equation}
        \dyad{\Phi({\bf v},t)},
    \end{equation}
       but the experimental detection is clearly tougher: even at fixed kinetic momentum with some error tolerance, there are not so many electrons to detect.
       We term this case {\it single-ionization ATI states}.
      \item Alternatively, we can condition on all ATI electrons, i.e., consider the reduced density matrix integrated over all outgoing momenta,
      \begin{equation}
      \int \dd^3 {\bf v}\dyad{\Phi({\bf v},t)}.
      \end{equation}
      
      Calculations and theoretical description is then more complex, but detection is easier.
\end{enumerate}

In the following, we explicitly elaborate on the calculations that lead to the states presented in the main text when considering this two different scenarios.

\subsection{Analysis for single-ionizaton ATI states}
The state shown in Eq.~\eqref{eq:ATI:SingleState:} is a superposition of the different coherent shifts generated during the ionization and acceleration processes, each of them multiplied by the matrix element $\meld^*_H({\bf v},t')$, which determines its correlation with the electron's state, associating each shift with the probability amplitude of having a transition from the ground state to the continuum state $\ket{{\bf v}}$. However, this state only considers transitions to a particular continuum state. 

In this subsection, we are going to consider single-ionization phenomena, i.e., laser ionization phenomena at a given kinetic momentum energy, corresponding to outgoing velocity $\textbf{v}$, so that the final ATI quantum state is indeed well characterized by the pure state given in Eq.~\eqref{eq:ATI:SingleState:}. Therefore, the obtained results would correspond to an experimental setting where we are able to measure the kinetic energy and direction of the generated photoelectrons, and discard the results whenever the measured kinetic energy and direction are different from those of $\textbf{v}$. This can be achieved by using the ATI photoelectron signal recorded by means of an time-of-flight electron spectrometer (see Fig.~\ref{FigSM2}).

In particular, and with the main purpose of obtaining analytical expressions, we will restrict this analysis to time intervals for which the applied strong field is constant, that is, $f(t') = 1$ in Eq.~\eqref{E:field:operator} for $t' \in [t_0, t]$. This implies that the amount of photons absorbed every half-cycle of the field would be the same, in opposition to Fig.~\ref{Fig1} (b) in the main text where the absorption varies every half-cycle due to the modulation of the applied pulse. In practice, this would correspond to a situation where the laser source is a ``long'' IR pulse, meaning that we can find several cycles with almost the same peak strength on its central part. Thus, and as a first step, we will rewrite Eq.~\eqref{eq:ATI:SingleState:} as a sum of integrals defined for every half-cycle of the field
\begin{equation}\label{Single:v:split}
    \begin{aligned}
    \ket{\Phi(\textbf{v},t)}
        &= i\hbar \sum_{j=0}^{\mathcal{N}-1}
        \int^{t_{j+1}}_{t_j} \! \! \! \dd t'\
        \opE_Q(t') \cdot \meld^*_H({\bf v},t')
        \\& \hspace{2cm}\times
        \ket{\delta\alpha(t')}
        \bigotimes_{q=2}^\text{cutoff}
        \ket{\beta_q(t')},
    \end{aligned}
\end{equation}
where $\mathcal{N}$ is the total number of half-cycles, and we identify $t_\mathcal{N} = t$. Note that the conditioning over a single value of direction and kinetic momentum ${\bf v}$ leads to an entangled state between all the modes participating in the process. Hereupon, and in order to study the final state obtained for the IR, we will assume that during the ATI process the harmonic coherent-state amplitudes $\beta_q$ stay very close to the vacuum. Thus, if under this assumption we project Eq.~\eqref{Single:v:split} over the vacuum state for the harmonics, we can approximate our state by
\begin{equation}\label{Single:v:split:fund}
    \begin{aligned}
    \ket{\Tilde{\Phi}(\textbf{v},t)}
        \approx i\hbar \sum_{j=0}^{\mathcal{N}-1}
        \int^{t_{j+1}}_{t_j} \! \! \! \dd t'\
        &\opE_{L}(t') \cdot \meld^*_H({\bf v},t')
        \ket{\delta \alpha(t')},
    \end{aligned}
\end{equation}
where $\ket{\Tilde{\Phi}(t)} = \bra{0_q}\bigotimes_q\ket{\Phi({\bf v},t)}$ and $\opE_L$ is the electric field operator acting over the fundamental mode, i.e., the first term of Eq.~\eqref{E:op:time:}.

Furthermore, under the ``long'' IR pulse considerations, the amount of photons absorbed every half-cycle is the same, that is, $\delta\alpha(t_{j+1}) - \delta \alpha(t_j) = \Delta$. This motivates us to consider a discretization of the values of $\delta \alpha(t)$ appearing on each term of the sum in Eq.~\eqref{Single:v:split:fund}, such that the value of $\delta \alpha(t)$ in each integral term adopts the value of the coherent-state obtained at the end of the cycle, that is, 
\begin{equation}\label{Single:ATI:discr}
    \ket{\Tilde{\Phi}(\textbf{v},t)}
        \approx i\hbar \sum_{j=0}^{\mathcal{N}-1}
        \int^{t_{j+1}}_{t_j} \! \! \! \! \! \dd t'\
        \opE_L(t') \cdot \meld^*_H({\bf v},t')
        \ket{(j+1)\Delta}.
\end{equation}

Of course, this approximation is not always valid. One has to guarantee that two consecutive states $\ket{j \Delta}$ and $\ket{(j+1)\Delta}$ are comparable to each other. Otherwise, smaller steps have to be considered in the discretization, which may not allow us to write the shift $\Delta$ as a time-independent quantity. A natural way of establishing such a comparison is in terms of the overlap between these two states, i.e.,
\begin{equation}\label{Constraint:Delta}
    \braket{j\Delta}{(j+1)\Delta}
        = \exp[-\dfrac{\lvert\Delta\rvert^2}{2}].
\end{equation}

Thus, we will restrict to values of $\lvert\Delta\rvert < 0.95$, for which the overlap between these two coherent-states is bigger than $1 - e^{-1}$. Under these considerations, the state obtained in Eq.~\eqref{Single:ATI:discr} is given as a superposition of different coherent-states, where each of them is affected by the electric field operator evaluated at time $t'$. Apart from this, one of the main differences of this state with respect to the one obtained through HHG, in Eq.~\eqref{eq:HHG:catstate:}, is that in the former more than two coherent-states intervene in the final superposition, depending on the number of half-cycles $\mathcal{N}$.

In Eq.~\eqref{Single:ATI:discr}, each of these coherent-states is weighted by the quantum optical version of the ATI spectrum taken at every half-cycle of the field. This can be seen more clearly if, assuming a linearly polarized field, we substitute Eq.~\eqref{E:field:operator} with the considered approximations in Eq.~\eqref{Single:ATI:discr}
\begin{equation}
    \begin{aligned}
    \ket{\phi(\textbf{v},t)}
        &\approx \hbar {\bf g}(\omega_L) \sum_{j=0}^{\mathcal{N}-1}
        \Bigg( 
            \int^{t_{j+1}}_{t_j} \! \! \! \! \! \dd t'\
                \meld^*_H({\bf v},t')e^{i\omega t'} \hat{a}
                \\& \hspace{2cm}
                -\meld^*_H({\bf v},t')e^{-i\omega t'} \hat{a}^\dagger
        \Bigg) \ket{(j+1)\Delta},
    \end{aligned}
\end{equation}
where
\begin{equation}
    \begin{aligned}
    \meld^*_H({\bf v},t)
        &= \mel{\psi_\text{sc}(t)}{\text{e}\hat{X}U_\text{sc}(t)}{{\bf v}}\\
        &= \mel{\psi_\text{sc}(t)}{\text{e}\hat{X}U_\text{sc}(t)}{{\bf p} -\frac{e}{c}\boldsymbol{A}_L(t_0)}.
    \end{aligned}
\end{equation}

In this last expression $\hat{X}$ is the position coordinate operator affecting the electron, $\hat{U}_\text{sc}(t)$ is the time evolution operator of the semiclassical Hamiltonian appearing in Eq.~\eqref{Sc:Sch:eq}, and $\ket{\psi_\text{sc}(t)}=U_\text{sc}(t)\ket{\text{g}}$ is the ground state of the electron evolved with the previous propagator. Furthermore, we have conditioned over kinetic energies that satisfy ${\bf v} = {\bf p} - (e/c)\boldsymbol{A}_L(t_0)$. Under the strong field assumptions, we can write the previous matrix element as
\begin{equation}
    \begin{aligned}
    \meld^*_H({\bf v},t) 
    &= 
    \mel{\psi_\text{sc}(t)}{\text{e}\hat{X}}{{\bf p} -\frac{e}{c}\boldsymbol{A}_L(t)}
    \\&\hspace{2cm}\times
    e^{-i(S({\bf p},t,t_0) - I_p(t-t_0))},
    \end{aligned}
\end{equation}
with $S({\bf p},t,t_0)$ the semiclassical action given in Eq.~\eqref{Semiclass:action}. By expanding this expression using the form of $\ket{\psi_\text{sc}(t)}$ given by the semiclassical analysis \cite{Lewenstein1994}, one can see that this term can be written as the sum of two terms characterizing direct ionization phenomena and rescattering processes \cite{Milosevic2006}. In our case, we are only interested in direct ionization processes, so we restrict our calculations to values of the electron kinetic energy lower than $2U_p$, with $U_p$ the ponderomotive potential. Thus, we write this matrix element as
\begin{equation}\label{Single:ATI:dipole}
    \meld^*_H({\bf v},t) 
        \approx \mel{\text{g}}{\text{e}\hat{X}}{{\bf p} -\frac{e}{c}\boldsymbol{A}_L(t)}e^{-i(S({\bf p},t,t_0) - I_p t)}.
\end{equation}

Now, we explicitly compute the expression for the Wigner function of the state in Eq.~\eqref{Single:ATI:discr}. With that purpose, let us first define the quantities $A_{j}$ and $B_j$ as
\begin{equation}
    \begin{aligned}
    &A_j 
        = \hbar {\bf g}(\omega_L)
          \int^{t_{j+1}}_{t_j} \dd t'
            \meld^*_H({\bf v},t) e^{i\omega t}\\
    &B_j
        = \hbar {\bf g}(\omega_L)
          \int^{t_{j+1}}_{t_j} \dd t'
            \meld^*_H({\bf v},t) e^{-i\omega t},
    \end{aligned}
\end{equation}
where $\meld^*_H({\bf v},t)$ is given as in Eq.~\eqref{Single:ATI:dipole}, such that the state in Eq.~\eqref{Single:ATI:discr} can be written as
\begin{equation}
    \ket{\Tilde{\Phi}(\textbf{v},t)}
        = i \sum_{j=0}^{\mathcal{N}-1}
            \big(A_j \hat{a} - B_j \hat{a}^\dagger)
            \ket{(j+1)\Delta}.
\end{equation}

Introducing here the definition of the photonic quadrature operators, $\hat{x}_L$ and $\hat{p}_L$ given in the main text, we can rewrite the previous state as
\begin{equation}
    \ket{\Tilde{\Phi}(\textbf{v},t)}
        = i\sum_{j=0}^{\mathcal{N}-1}
            \big(C_j^{(-)} \hat{x}_L + iC_j^{(+)} \hat{p}_L)
            \ket{(j+1)\Delta},
\end{equation}
where $C_j^{\pm} = (1/\sqrt{2})(A_j \pm B_j)$. Thus, for computing the Wigner function by means of 
\begin{equation}\label{ATI:Wigner:equal:shift}
    W(x,p) = \dfrac{1}{\pi \hbar}
        \int^{\infty}_{-\infty}
            \mel{x+y}{\Tilde{\rho}_\text{ATI-IR}}{x-y}
            e^{-i2py/\hbar},
\end{equation}
we first give an expression for the matrix element of $\rho = \dyad{\Tilde{\Phi}(\textbf{v},t)}$ between two different position states $\ket{x\pm y}$
\begin{equation}\label{Single:ATI:Wig:ME}
    \begin{aligned}
    \mel{x+y}{\rho}{x-y}
        &= \Big[
                \sum_{j=0}^{\mathcal{N}-1}
                C_j^{(+)}\mel{x+y}{\hat{x}_L}{(j+1)\Delta}
                 \\ &
                \hspace{1.5cm} +iC_j^{(-)}\mel{x+y}{\hat{p}_L}{(j+1)\Delta}
            \Big]
            \\
            & \times
            \Big[
                \sum_{k=0}^{\mathcal{N}-1}
             C_k^{(+)*}\mel{(k+1)\Delta}{\hat{x}_L}{x-y}
             \\ &
                \hspace{1.5cm}
                -iC_k^{(-)*}\mel{(k+1)\Delta}{\hat{p}_L}{x-y}
            \Big],
    \end{aligned}
\end{equation}
with
\begin{equation}
    \begin{aligned}
    &\mel{x+y}{\hat{x}_L}{(j+1)\Delta}
        = (x+y)G_{+,j}\\
    &\mel{x+y}{\hat{p}_L}{(j+1)\Delta}
        = -i \pdv{G_{+,j}}{(x+y)},
    \end{aligned}
\end{equation}
where the functions $G_{\pm,j} = \braket{x\pm y}{(j+1)\Delta}$ are given by
\begin{equation}
    \braket{x}{\alpha}
        = \dfrac{1}{\pi^{1/4}}
            \exp[
                 -\dfrac{(x-\sqrt{2}\Re(\alpha))^2}{2}
                 + i x \sqrt{2} \Im(\alpha)
                ].
\end{equation}

With all this, the matrix element in Eq.~\eqref{Single:ATI:Wig:ME} reads
\begin{equation}
    \begin{aligned}
    \mel{x+y}{\rho}{x-y}
        &= \sum_{j,k}^{\mathcal{N}-1}
            \bigg[
                C_{j}^{(-)}C_{k}^{(-)*}
                    (x^2-y^2)G_{+,j}G^*_{-,k}
                    \\
                &+ C_{j}^{(+)}C_{k}^{(+)*}
                    \pdv{G_{+,j}}{(x+y)}\pdv{G^*_{-,k}}{(x-y)}\\
                &+C_{j}^{(-)}C_{k}^{(+)*}
                    (x+y)G_{+,j}\pdv{G^*_{-,k}}{(x-y)}
                       \\
                &+
                    C_{j}^{(+)}C_{k}^{(-)*}
                 (x-y)\pdv{G_{+,j}}{(x+y)}G^*_{-,k}
            \bigg],
    \end{aligned}
\end{equation}
and, thus, the Wigner function can be computed by introducing this expression for the matrix element inside Eq.~\eqref{ATI:Wigner:equal:shift}. Note that this expression will only contain derivatives involving gaussian functions, so it can computed analytically. In particular, we have performed these calculations in atomic units ($\hbar = 1, e^2 = 1, m_e = 1$ and $k_c = 1/4\pi\varepsilon_0 = 1$). In particular, we considered the ionization potential of an hydrogen atom $I_p = 0.5$ a.u., the frequency for the fundamental mode $\omega = 0.057$ a.u., and the amplitude of the electromagnetic field $E_L = 0.053$ a.u.

\subsection{ATI state conditioned over all possible outgoing momenta}
The density matrix that characterizes the total IR ATI state involving all the possible momenta for the generated photoelectrons is  
\begin{equation}
	\begin{aligned}
	\rho_\text{ATI}
		&= \int \dd {\bf v}  \dyad{\Phi({\bf v},t)}{\Phi({\bf v},t)}\\
		&= \int \dd {\bf v} \int_{t_0}^t\dd t' \int_{t_0}^t \dd t''
		  \ \opE_Q(t') \cdot \meld^*_H({\bf v},t')
		  \\ & \hspace{1cm}\times
		  		\dyad{\Phi(t')}{\Phi(t'')}
		  		    \meld_H({\bf v},t'') \cdot \opE_Q(t''),
	\end{aligned}
\end{equation}
which, taking into account the SFA version of the identity, i.e.,
\begin{equation}
    \mathbbm{1} \approx
        \dyad{\text{g}} + \int \dd \vb{v} \dyad{\vb{v}}
\end{equation}
and considering for simplicity a linearly polarized light, can be rewritten as
\begin{equation}\label{Simplified:state}
    \begin{aligned}
    \rho_\text{ATI}
        &= \int^t_{t_0} \dd t' \int^t_{t_0} \dd t'' \
            \hat{E}_Q(t')
                \dyad{\Phi(t')}{\Phi(t'')}
            \hat{E}_Q(t'')
            \\&\hspace{1.25cm}\times
            \big[\langle \hat{d}_H(t') \hat{d}_H(t'') \rangle
                - d_H(t') d_H(t'')\big],
    \end{aligned}
\end{equation}
where the term between brackets contains the difference between the correlation of the dipole operator at times $t'$ and $t''$, and the product of the mean values of such operators at the corresponding times, both terms evaluated with respect to the ground state of the system. For other possible field polarizations, the expression adopts the same form but we would have to consider contributions coming from the different polarization terms for the term between brackets. Obviously, while measurement conditioned on all electrons should be easier, the theoretical analysis is tougher as it requires evaluation of the two-time correlation functions of the dipole moment. This can be done, in principle using SFA or even TDSE, but leads to much more complicated expressions, which will be analysed elsewhere \cite{PhilippQED}.

In order to gain intuition about IR ATI state obtained from Eq.~\eqref{Simplified:state}, we are going to work within the same approximations that lead to Eq.~\eqref{Single:v:split:fund}, and considering the simplifying assumption that all the generated coherent shifts are identical and time-independent. In general this is not true and, as discussed in Fig.~\ref{Fig1} in the main text, the coherent shift is continuously increasing along the pulse. However, for single photon ionization processes one may expect this shift to be very small and, in some sense, indistinguishable from all the other values it can take along the whole pulse duration. Therefore, under this consideration the ATI state reads
\begin{equation}
\begin{aligned}
    \Tilde{\rho}_\text{ATI-IR}
        &= \int^t_{t_0} \dd t' \int^t_{t_0} \dd t'' \
            \hat{E}_L(t')
                \dyad{\delta \alpha}
            \hat{E}_L(t'')
            \\&\hspace{1.25cm}\times
            K(t',t'') e^{i\varphi(t')}e^{-i\varphi(t'')},
    \end{aligned}
\end{equation}
where $K(t',t'') = \langle \hat{d}_H(t') \hat{d}_H(t'') \rangle- d_H(t') d_H(t'')$, and the exponential terms are the factors coming from the BCH formula, which we have to explicitly consider as they cannot be factorized now. Furthermore, if we introduce here the definition of part of electric field operator that acts over the fundamental mode (first term in Eq.~\eqref{E:field:operator}), we get 
\begin{equation}
    \begin{aligned}
    \Tilde{\rho}_\text{ATI-IR}
    &= \hbar^2 \lvert \vb{g}(\omega_L)\cdot \boldsymbol{\epsilon}_{\mu,L}\rvert^2
    \int^t_{t_0} \dd t' \int^t_{t_0} \dd t'' K(t',t'')
    \\&\hspace{1.25cm}\times
    \Big[
        \hat{a}^\dagger \dyad{\delta \alpha} \hat{a}^\dagger
            e^{i\omega_L(t'+t'')}
        \\&\hspace{1.75cm}
        + \hat{a} \dyad{\delta \alpha} \hat{a}
            e^{-i\omega_L(t'+t'')}
            \\&\hspace{1.75cm}
        - \hat{a}^\dagger \dyad{\delta \alpha} \hat{a}
            e^{i\omega_L(t'-t'')}
        \\&\hspace{1.75cm}
        - \hat{a} \dyad{\delta \alpha} \hat{a}^\dagger
            e^{-i\omega_L(t'-t'')}
    \Big].
    \end{aligned}
\end{equation}

Thus, one of the main advantages of the previous approximation is that the temporal temporal part only affects the coefficients of the obtained mixed state. This allow us to write Eq.~\eqref{Simplified:state} as
\begin{equation}\label{Simplified:state:expanded}
    \begin{aligned}
    \Tilde{\rho}_\text{ATI-IR} 
        &= -i\hbar \lvert \vb{g}(\omega_L)\cdot
            \boldsymbol{\epsilon}_{\mu,L}\rvert
            \\ &\quad\times
            \Big[
                I_1(t) 
                    \hat{a}^\dagger \dyad{\delta \alpha} \hat{a}^\dagger
                + I_2(t)
                    \hat{a} \dyad{\delta \alpha} \hat{a}\\
                &\hspace{0.5cm}
                - I_3(t)
                    \hat{a}^\dagger \dyad{\delta \alpha} \hat{a}
                - I_4(t)
                    \hat{a} \dyad{\delta \alpha} \hat{a}^\dagger
            \Big],
    \end{aligned}
\end{equation}
where we have defined
\begin{align}
    &I_1(t) =
        \int^t_{t_0} \dd t' \int^t_{t_0} \dd t''
            \bar{K}(t',t'') e^{i\omega_L(t'+t'')},\\
    &I_2(t) = 
        \int^t_{t_0} \dd t' \int^t_{t_0} \dd t''
            \bar{K}(t',t'') e^{-i\omega_L(t'+t'')},\\
    &I_3(t) = 
        \int^t_{t_0} \dd t' \int^t_{t_0} \dd t''
            \bar{K}(t',t'') e^{i\omega_L(t'-t'')},\\
    &I_4(t) = 
        \int^t_{t_0} \dd t' \int^t_{t_0} \dd t''
            \bar{K}(t',t'') e^{-i\omega_L(t'-t'')},
\end{align}
with $\bar{K}(t',t'') = K(t',t'')e^{i\varphi(t')}e^{-i\varphi(t'')}$.

The dipole correlator $K(t',t'')$ is a difficult to compute quantity, since it is not a quantity that we can be obtained directly from the numerical implementations of the TDSE, neither from a SFA analysis. Thus, the approach we consider here in order to gain intuition about what to expect of the obtained Wigner functions, is to look for some relations between the $I_i(t)$ coefficients so that we can bring Eq.~\eqref{Simplified:state:expanded} to a very simplified form, and then study different limits regarding the coefficients. First of all, we note that the $\bar{K}(t',t'')$ satisfies $\bar{K}(t',t'') = \bar{K}^*(t'',t')$, which allow us to conclude after some algebraic operations that $I_1(t) = I^*_2(t)$ and that $I_3(t)$ and $I_4(t)$ are real functions. 

The above relations allow us to further simplify the final form of $\Tilde{\rho}_\text{ATI-IR}$, and provides us with the final form we use for the Wigner function computation
\begin{equation}\label{Simplified:state:expanded2}
    \begin{aligned}
    \Tilde{\rho}_\text{ATI-IR} 
        &= -i\hbar \lvert \vb{g}(\omega_L)\cdot
            \boldsymbol{\epsilon}_{\mu,L}\rvert
            \\&\quad\times
            \Big[
                I_1(t) 
                    \hat{a}^\dagger \dyad{\delta \alpha} \hat{a}^\dagger
                + I^*_1(t)
                    \hat{a} \dyad{\delta \alpha} \hat{a}\\
                &\hspace{1cm}
                - I_3(t)
                    \hat{a}^\dagger \dyad{\delta \alpha} \hat{a}
                - I_4(t)
                    \hat{a} \dyad{\delta \alpha} \hat{a}^\dagger
            \Big].
    \end{aligned}
\end{equation}

Then, using the definition of the Wigner function given in \cite{Royer1977}, we get for our state
\begin{equation}
    \begin{aligned}
    W(\beta)
            &= \dfrac{2}{\pi N}
            e^{-\tfrac12\lvert 2\beta-\delta\alpha\rvert}
            \\&\quad\times
            \Big[
                I_1(t) 
                    \delta\alpha(2\beta-\delta\alpha)
                + I^*_1(t)
                    \delta\alpha^*(2\beta-\delta\alpha)^*\\
                &\hspace{1cm}
                - I_3(t)
                    \lvert\delta\alpha\rvert
                - I_4(t)
                    (\lvert 2\beta-\delta\alpha\rvert -1)
            \Big],
    \end{aligned}
\end{equation}
where $N$ is a normalization constant. As it was mentioned before, the computation of the $K(t,t')$ function is not trivial at all, and in the strong-field community it is common to approximate the absolute value of the Fourier transform of the dipole-dipole correlator with the absolute value of the Fouier transform given by the dipole, i.e., the fundamental component of the HHG spectrum (c.f. \cite{Sand2000}). In our case, and to gain insight about the form of the final Wigner function, we neglect the effect of the exponentials with respect to $(t'+t'')$ as we expect their contribution to be lower than the ones provided by $(t'-t'')$, as their oscillation is faster. With this, one can check that for different values of the weights provided by the integrals $I_3(t)$ and $I_4(t)$, the final Wigner function presents a similar behavior to the one obtained in HHG. Because of the form of the considered quantum state, this is something we should expect since $\hat{E}_L(t) \ket{\delta \alpha} \propto \ket{\delta \alpha}$ when $\delta\alpha$ adopts very large values.

\begin{figure*}
		\includegraphics[width=0.8 \textwidth]{FigSM2.pdf}
		\caption{Operation principle of the experimental approach. \textbf{(a)} Experimental setup. BS$_{1}$: IR beam separator. $\ket{\alpha_L}$: coherent-state of IR beam passing through BS$_{1}$. $\ket{\alpha_r}$: coherent-state IR beam reflected by BS$_{1}$. M: IR plane mirrors. L$_{1,2}$: Lens. HS: harmonic separator which reflects the high harmonics and lets the IR beam pass through. HH: High harmonics. BS$_{2,3}$: IR beam separator and splitter, respectively. PD, PD$_{0}$, PD$_{out}$, PD$_{HH}$: IR and HH photodetectors. TOF e-Spec.: $\mu$--metal shielded time of flight spectrometer that could be used for the measurement of the ATI electrons. The voltage (-V) can serve for the energy selection of the electrons reaching the TOF detector. $i_{\phi}$, $i_{out}$, $i_{0}$, $i_{HH}$, are the photocurrent values recorded for each laser shot. These were used by the QS in order to condition the $\ket{\alpha_L+\delta\alpha_L}$ state on the HHG process. $i_{e}$ is the signal of the TOF spectrometer that could be used by QS for conditioning on ATI process. Just before PD$_{HH}$ a 150 nm thick aluminum filter was placed (not shown) in order to select the  harmonics with $q \geq 11$. IR$_{0}$: IR beam used to measure the shot energy of the driving field. $\lambda/2$: Half-IR-wave plates. A: Apertures. F, F$_{in}$: Neutral density filters of approximately the same transmission. F$_{r}$: Neutral density filters used to control the energy of the reference coherent-state of the laser field $E_{r}$.  $\ket{\alpha+\delta\alpha}$: IR state after the attenuation. All signals were recorded by a high dynamics range boxcar integrator and saved/analyzed by computer (PC) software. $\ket{\Phi_\text{post}}$ is the quantum state of field entering the balance detector after conditioning on HHG, and $E_{in}$ is the corresponding electric field. $\ket{\alpha_r}$ is the reference coherent-state of the laser used by the QT method, and $E_{r}$ is the corresponding electric field. $\varphi$: The controllable phase shift introduced in the reference beam. \textbf{(b)} HHG spectra measured for two different xenon gas densities in the interaction region. The blue and green lines show the harmonics recorded at high and low gas densities that have used for the generation of the optical cat and kitten states shown in Figs.~\ref{Fig7}b and ~\ref{Fig7}c of the main text of the manuscript. The harmonic signal at low gas densities is about 25 times lower than the harmonic signal recorded at high gas densities. \textbf{(c)} Probability of absorbing IR photons towards the harmonic generation (red line). The multi--peak structure reflects the spectrum of the emitted harmonics as is described in \ref{SM4} and refs \cite{Lewenstein1, Tsatrafyllis1, Tsatrafyllis2}. The black dashed-dot curve is the best fit of an analytical function given by the sum of a sequence of gaussian functions. The black shaded area shows the background distribution resulted by fitting a gaussian function on the data (black squares) obtained by subtracting the minima of the raw data from the minima of the black dashed-dot fit function. The Inset shows the joint XUV--vs--IR photon number distribution using the signal of $i_{HH}$ ($S_{PD_{HH}}$) and $i_{out}$ ($S_{PD_{out}}$) (gray points). The red points show the selected points along the anti--correlation diagonal. The distribution was created by keeping the energy stability of the driving field at the level of $\approx 1$\%, and after subtracting the electronic noise from each laser shot.} 
		\label{FigSM2}
\end{figure*}

\begin{figure*}[ht!]
	\includegraphics[width=0.9 \textwidth]{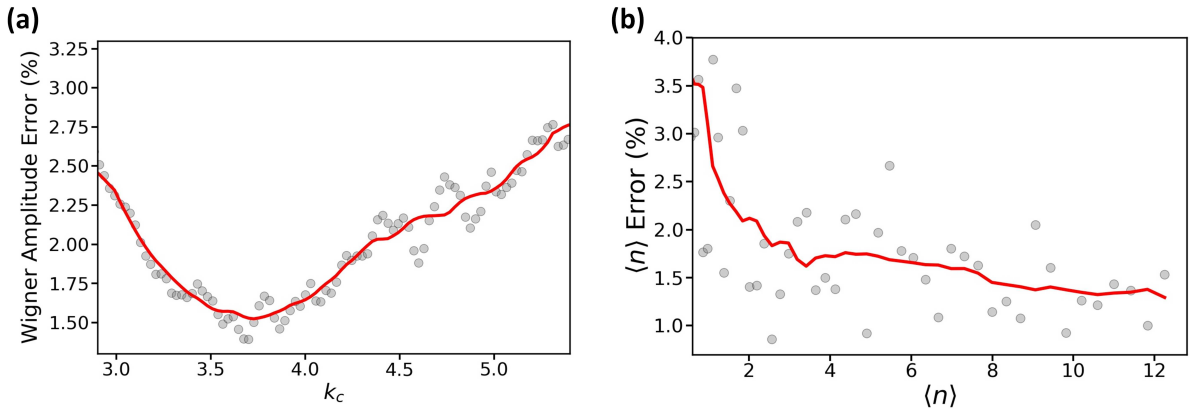}
	\caption{Error analysis of the reconstructed Wigner function and the photon number. \textbf{(a)} Dependence of the error of the Winger function reconstructed by the experimental data on $k_{c}$. \textbf{(b)} Dependence of accuracy of measuring the photon number on the mean photon number of the light state. In both graphs, the red solid line is a 15 points running average of the data (gray points).} 
	\label{FigSM3}
\end{figure*}

\section{Operation principle of the experimental approach}\label{SM6}

An optical layout of the system is shown in Figure~\ref{FigSM2}a. Although the system can be implemented for conditioning on HHG and/or ATI processes, here we will show its applicability using the HHG process induced by the interaction of the fundamental driving field with Xe gas. The approach has been also discussed in ref.~\cite{Lewenstein1}. The experiment was performed using a linearly polarized $\approx$ 35 fs Ti:Sapphire laser pulse of $\lambda\approx$ 800 nm carrier wavelength and an interferometer. The whole system was operating at 0.5 kHz repetition rate. The IR laser beam was separated into the branches of the interferometer by a beam separator BS$_1$. The reflected by the BS$_1$ IR beam (in the $2^\text{nd}$ branch of the interferometer) serves as a reference beam of the quantum tomography (QT) method and for measuring (by means of IR photodiode PD$_{0}$) the shot-to-shot energy fluctuations of the driving field. In the 1$^\text{st}$ branch of the interferometer, the IR beam was focused by means of a 15 cm focal length lens (L$_{1}$) into a xenon pulsed gas jet, where the HHG process takes place. In the present experiment, the optimum intensity of the IR pulse in the interaction region resulting to a maximum harmonic order was $\approx 8 \times 10^{13}$ W/cm$^2$, while the maximum harmonic yield was observed for a gas density in the order $\sim 10^{18}$ atoms/cm$^3$. The generated harmonics, after a reflection by a multilayer infrared-antireflection coating plane mirror (HS) placed at grazing incidence angle, was passing through a 150 nm thick aluminum filter, which selects all the harmonics with $q \geq 11$ Figure~\ref{FigSM2}b. The photon number of the XUV radiation was measured by means of a calibrated XUV detector PD$_{HH}$. A portion of the IR field exiting the xenon gas was reflected by the IR beam separator BS$_2$ towards IR photodiode PD$_{out}$ (operating in the linear regiem) placed after a lens (used to collect the photons on the surface of the diode, not shown in Fig.~\ref{FigSM2}a) and a neutral density filter (F) which significantly reduces the photon number and ensures the avoidably of saturation effects. The photocurrent signals $i_{HH}$, $i_{0}$, $i_{out}$ of PD$_{HH}$, PD$_{0}$ and PD$_{out}$ were used by the quantum spectrometer (QS) to disentangle the high harmonic generation process from all other processes induced by the interaction. The IR field after BS$_2$ was collimated by a plano-convex lens (L$_{2}$) while the mean photon number of the IR field (E$_{in}$), before reaching the balanced detector of the QT, was reduced (by means of neutral density filters F$_{in}$) to the level of few photons per pulse, with the QS to select, for each laser shot, only the IR photons related to the HHG. The QS approach \cite{Tsatrafyllis1, Tsatrafyllis2} relies on shot-to-shot correlation between the photon number of the generated harmonics (integrated signal of $q \geq 11$) and the IR field exiting the medium (gray points in the inset of Figure~\ref{FigSM2}c). The conditioning to HHG is achieved by selecting only the shots that provide signal along the anti-correlation diagonal of the joint distribution (red points in the inset Figure~\ref{FigSM2}c). By selecting these points, we collect only the shots that are relevant to the harmonic emission and we remove the unwanted background associated with all processes irrelevant to the harmonic generation. In this way, we obtain the probability of absorbing IR photons towards the harmonic generation (red line in Figure~\ref{FigSM2}c). The IR absorption probability distribution consists on a multi-peak structure which corresponds to the harmonic order \cite{Tsatrafyllis1, Tsatrafyllis2}. The black line in Figure~\ref{FigSM2}c shows the remaining background distribution which needs, and has been subtracted from the data, as is related only with the ability of the present QS experimental apparatus to remove all the shots associated with processes irrelevant to the HHG process (for details see refs. \cite{Tsatrafyllis1, Lewenstein1, Lamprou2021}).

The E$_{in}$ field was spatiotemporally overlapped on a beam splitter (BS$_{2}$) with an unaffected by the interaction local oscillator laser field (E$_{r}$) coming from the 2$^\text{nd}$ branch of the interferometer which consists of a piezo-based delay stage that introduces a controllable delay $\Delta\tau$ (phase shift $\varphi$) between the E$_{r}$ and E$_{in}$ fields. The outgoing from the BS$_{2}$ interfering fields were detected by the diodes (PD) of a high bandwidth (from DC to 350 MHz), high subtraction efficiency and high quantum efficiency, balanced amplified differential photodetector, which provides at each value of $\varphi$ the signal difference. The photocurrent difference $i_{\varphi}$, as well as the photocurrent values of the IR and HH detectors ($i_{out}$, $i_{0}$, $i_{HH}$) in the QS, were simultaneously recorded for each laser shot by a multichannel 16 bit high dynamic range boxcar integrator. For each shot the background electronic noise was recorded and subtracted by the corresponding photocurrent signal by placing a second time-gate in the boxcar integrator in times significantly delayed compared to the arrival times of the photon signals. Setting the delay stage around $\Delta\tau\approx 0$, the characterization of the quantum state of light was achieved by recording for each shot the value of $i_{\varphi}$ as a function of $\varphi$, by moving the piezo from $\varphi \approx 0$ to $\varphi \approx \pi$. The homodyne data was scaled according to the measured vacuum state quadrature noise. 

\section{Reconstruction of the Wigner function}\label{SM7}
The values of the photocurrent difference $i_{\varphi}$ are directly proportional to the measurement of the electric field operator $\hat{E}_{in} (\varphi) \propto \hat{x}_{\varphi}=\cos(\varphi) \hat{x}+ \sin(\varphi) \hat{p}$, and have been used for the reconstruction of the Wigner function. When the xenon gas jet and the QS was switched on the homodyne detection system provides the measurement $\hat{x}_{\varphi}$ only when IR field exiting the atomic medium is conditioned on the HHG, leading to the characterization of the light state $\ket{\Phi_\text{post}}=\ket{\alpha_L+\delta\alpha_L}- \xi \ket{\alpha_L}$. Repeated measurements of $\hat{x}_{\varphi}$  at each $\varphi$ provides the probability distribution $P_{\varphi}(x_{\varphi})=\langle{x_{\varphi}}|{\hat{\rho}}|{x_{\varphi}}\rangle$ of its eigenvalues $x_{\varphi}$ (where $\hat\rho\equiv\dyad{\Phi_\text{post}}$ is the density operator of the light state and $|{x_{\varphi}}\rangle$ the eigenstate with eigenvalue $x_{\varphi}$). For each data set in the range of $0 < \varphi < \pi $ around $\Delta\tau \approx 0$, the Wigner function was reconstructed by means of the inverse Radon transformation implemented via the standard filtered back-projection algorithm \cite{QT1, QT2}. The algorithm used to reconstruct the Wigner functions was applied directly to the quadrature values $x_{\varphi,k}$, where $k$ is the index of each value, using the formula\cite{QT1, QT2} $W(x,p)\simeq \frac{1}{2\pi^{2}N} \sum_{k=1}^{N}K(x\cdot \cos(\varphi_{k})+p\cdot \sin(\varphi_{k})-x_{\varphi, k})$. $K(z)=\frac{1}{2} \int_{-\infty}^{\infty}|\xi|\exp(i\xi z)\,d\xi$ is called integration kernel with $z=x\cdot \cos(\varphi_{k})+p\cdot \sin(\varphi_{k})-x_{\varphi,k}$. The numerical implementation of the integration kernel requires the replacement of the infinite integration limits with a finite cutoff frequency $k_{c}$. In order to reduce the numerical artifacts (rapid oscillations) and allow the details of the Wigner function to be resolved, the value of $k_{c}$ was set to $\approx$ 3.7 for all measurements presented here. An estimation of the error of the reconstructed $W(x,p)$ has been obtained by comparing (subtracting) the ideal Wigner function of a coherent-state from the Wigner function of a coherent-state reconstructed by the experimental data. The deviation from the ideal case provides an error of $\pm0.004$ in $W(x,p)$. The accuracy of measuring the photon number was in the range of $\approx1.5\%$ to $\approx3.5\%$ of the mean, for high and low photon numbers, respectively. This was obtained following the aforementioned procedure using the density matrices $\rho_{nm}$ in Fock space $(n,m)$. The mean photon number was obtained by the diagonal elements $\rho_{nn}$ of the $\rho_{nm}$ and the relation $\langle{n}\rangle=\sum n\rho_{nn}$.

\section{Error analysis of the reconstructed Wigner function and the photon number}\label{SM8}

The numerical implementation of the integration kernel for the reconstruction of the Wigner function, requires the replacement of the infinite integration limits with a finite cutoff frequency $k_{c}$. In order to reduce the numerical artifacts (rapid oscillations) and allow the details of the Wigner function to be resolved, the value of $k_{c}$ was set to $\approx$ 3.7 for all measurements presented here. An estimation of the error of the reconstructed $W(x,p)$ has been obtained by comparing (subtracting) the ideal Wigner function of a coherent-state from the Wigner function of a coherent-state reconstructed by the experimental data. This is shown in Fig.~\ref{FigSM3}a as a function of $k_{c}$. The deviation from the ideal case provides an error $\approx1.5\%$ resulting an error of $\pm0.004$ in the $W(x,p)$ shown in the main text of the manuscript. This Figure also shows that used value of $k_{c}\approx$ 3.7 is indeed the optimum.

To obtain the accuracy of measuring the photon number, we have followed the aforementioned procedure for each light state shown in the main text of the manuscript, using the density matrices $\rho_{nm}$ in Fock space $(n,m)$. The mean photon number was obtained by the diagonal elements $\rho_{nn}$ of the $\rho_{nm}$ and the relation $\langle{n}\rangle=\sum n\rho_{nn}$. The results shown in Fig.~\ref{FigSM3}b have been obtained by calculating the mean photon number value ($\langle n_{rec} \rangle$) of a coherent-state numerically constructed using the number of data points recorded in the experiment. This value has been compared with the value resulting from the ideal theoretical case ($\langle n_{th} \rangle$) i.e. we obtain the ($\langle n \rangle$ Error (\%) $= |\langle n_{rec} \rangle-\langle n_{th} \rangle| / \langle n_{th} \rangle$). This procedure has been repeated for different photon number values of the coherent-state. It is found that the accuracy of measuring the photon number is in the range of $\approx1.5\%$ to $\approx3.5\%$ of the mean, for high and low photon numbers, respectively.

\section{\textit{Ab-initio} analysis of the decoherence due to the interaction with an environment}\label{SM9}
Here, we further extend our calculations to the interaction of the obtained HHG Schrödinger optical cat states with an environment. In particular, the model we consider is that of a beam splitter, where in one of the input modes we introduce the quantum state we want to study, and on the other an ancillary vacuum state which is traced out at the output. Thus, we can understand this ancillary mode as the part of the field which is absorbed by the environment. Although simple, this model has been proven to be exact when describing interactions with a Gaussian reservoir \cite{Leonhardt1993}, and we show here that describes the differences obtained between the theoretical and experimental Wigner functions.

\begin{figure}
    \centering
    \includegraphics[width = 1.\columnwidth]{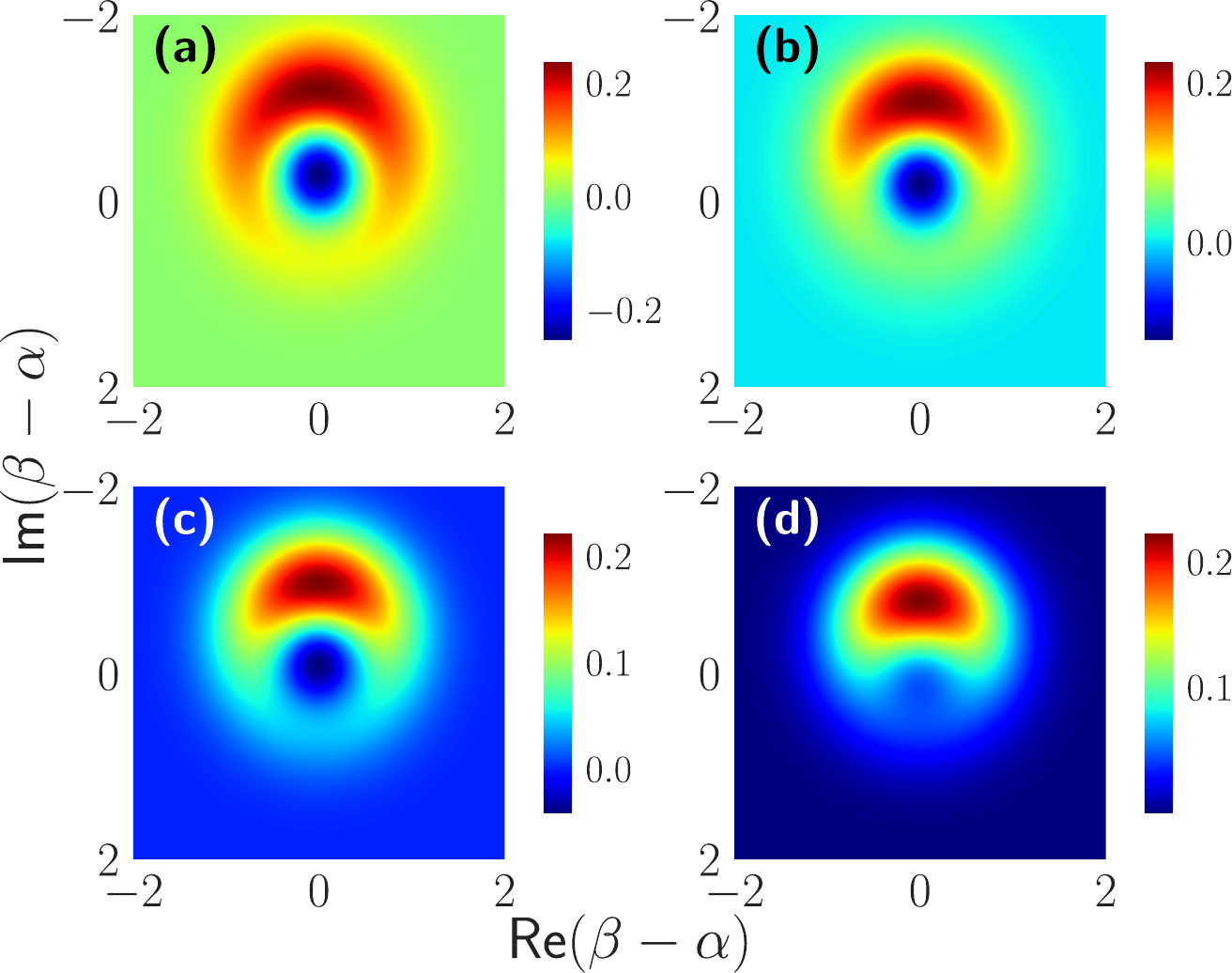}
    \caption{Wigner functions of the cat state after considering its interaction with the environment. Here, we consider $\delta \alpha = -0.8i$ and transmission efficiencies (a) $\eta=1.0$, (b) $\eta=0.75$, (c) $\eta=0.59$ and (d) $\eta=0.39$. The different axes characterize the different quadratures of the field, in particular $\Re[\beta-\alpha]\equiv x_L, \Im[\beta-\alpha]\equiv p_L$, with $x_L$, $p_L$ the values of the quadrature field operators $\hat{x}_L=(\hat{a}+\hat{a}^\dagger)/\sqrt{2}$ and $\hat{p}_L=(\hat{a}-\hat{a}^\dagger)/i\sqrt{2}$. }
    \label{FigSM4}
\end{figure}

According to this model, the state after the interaction with the environment is described by
\begin{equation}
    \Tilde{\rho}
        = \tr_{\text{anc}}
            \big(
                B(\theta) 
                    \dyad{\Phi_\text{post}}
                    \otimes \dyad{0_\text{anc}} 
                B(\theta)^\dagger
            \big),
\end{equation}
where $\tr_{\text{anc}}$ represents the partial trace over the ancillary mode, $B(\theta)\equiv \exp[\theta(\hat{a}\hat{a}^\dagger_\text{anc} - \hat{a}^\dagger\hat{a}_\text{anc})]$ is a unitary operator describing the beam splitter where $\hat{a}_\text{anc}$ ($\hat{a}_\text{anc}^\dagger$) is the annihilation (creation) operator acting over the ancillary modes, $\theta$ is a parameter related to the transmission efficiency $\eta$ by $\eta = \cos[2](\theta)$, and $\ket{\Phi_\text{post}}$ is the HHG optical cat state given in Eq.~\eqref{eq:HHG:catstate:}.

We find for the noise-affected state
\begin{equation}
    \begin{aligned}
    \Tilde{\rho} &= \dfrac{1}{N}
        \Big[
            \dyad{(\alpha + \delta\alpha)\cos(\theta)}
            \\ & \hspace{1cm}
            + \lvert\xi\rvert \dyad{\alpha\cos(\theta)}\\
            &\hspace{1cm}
            - \xi \Tilde{\xi} 
                \dyad{\alpha \cos(\theta)}{(\alpha+\delta\alpha)\cos(\theta)}
            \\ & \hspace{1cm}
            - \xi^* \Tilde{\xi}^*
                \dyad{(\alpha+\delta\alpha) \cos(\theta)}{\alpha\cos(\theta)}
        \Big],
    \end{aligned}
\end{equation}
where $\xi = \braket{\alpha}{\alpha + \delta \alpha}$, $\Tilde{\xi} = \braket{\alpha\sin(\theta)}{(\alpha + \delta \alpha)\sin(\theta)}$ and $N$ is the normalization factor. Using the definition for the Wigner function provided in \cite{Royer1977}, we find
\begin{equation}
    \begin{aligned}
    W(\beta) 
        &= \dfrac{2}{\pi N}
            \bigg[
                e^{-2\lvert\beta - (\alpha+\delta\alpha)\cos(\theta)\rvert}
                + \lvert \xi \rvert
                    e^{-2\lvert\beta - \alpha\cos(\theta)\rvert}
                \\
                &\hspace{0.5cm}
                -\Big(
                    \xi\Tilde{\xi}
                    e^{-i 2\Im(\beta)\delta\alpha\cos(\theta)}
                    +\xi^*\Tilde{\xi}^*
                    e^{i 2\Im(\beta)\delta\alpha\cos(\theta)}
                \Big)
                \\&\hspace{1.25cm}\times
                    e^{-\tfrac12\lvert 2\beta - (2\alpha + \delta \alpha)\cos(\theta)}
            \bigg],
    \end{aligned}
\end{equation}
and whose main features are shown in Fig.~\ref{FigSM4}. In these plots, we considered $\delta\alpha=-0.8i$ and decreasing values, from (a) to (d), of the transmission efficiency. As we can see, the Wigner distributions keep their shape while the negative regions become smaller. Evidently, in the case of zero transmissivity, we get a Gaussian distribution that is centered in the origin. These features describe very well the experimental observations, where the negative regions become very small compared to the theoretical values. However, in the experiment we also have the noise contributions coming from the measurement devices, which are not captured by this simple model. 

\bibliography{Paper.bib}{}
\end{document}